\documentclass[range]{ar2e}
\begin{document}

\input epsf.tex    
\input epsf.def   

\input psfig.sty

\jname{Ann. Rev. Nucl. Part. Sci.} \jyear{2000} \jvol{}
\ARinfo{1056-8700/97/0610-00}


\def\eq{\begin{equation}}
\def\eeq{\end{equation}}
\def\eqa{\begin{eqnarray}}
\def\eeqa{\end{eqnarray}}

\def\nn{\nonumber}
\def\veps{\varepsilon}
\def\pref#1{(\ref{#1})}

\def\Sca{{\cal A}}
\def\Scd{{\cal D}}
\def\Scf{{\cal F}}
\def\Scl{{\cal L}}
\def\Sco{{\cal O}}
\def\Scw{{\cal W}}

\def\bfj{{\bf j}}
\def\bfk{{\bf k}}
\def\bfp{{\bf p}}
\def\bfr{{\bf r}}

\def\Bfi{{\bf I}}
\def\Bfj{{\bf J}}
\def\Bfv{{\bf V}}

\def\ssb{{\scriptscriptstyle B}}
\def\sse{{\scriptscriptstyle E}}
\def\ssf{{\scriptscriptstyle F}}
\def\ssj{{\scriptscriptstyle J}}
\def\ssl{{\scriptscriptstyle L}}
\def\ssm{{\scriptscriptstyle M}}
\def\ssr{{\scriptscriptstyle R}}
\def\ssv{{\scriptscriptstyle V}}
\def\ssw{{\scriptscriptstyle W}}
\def\ssy{{\scriptscriptstyle Y}}

\def\Scabr{{\overline{\cal A}}}
\def\hbr{{\overline{h}}}
\def\chibr{\overline{\chi}}
\def\psibr{\overline{\psi}}
\def\nubr{\overline{\nu}}
\def\ol#1{\overline{#1}}

\def\G{\Gamma}
\def\msbar{\overline{MS}}

\def\eg{{\it e.g.}}
\def\ie{{\it i.e.}}
\def\etc{{\it etc.}}

\def\oneloop{{\rm 1-loop}}
\def\lowe{{\rm l.e.}}
\def\hie{{\rm h.e.}}
\def\leff{\Scl_{\rm eff}}
\def\seff{S_{\rm eff}}
\def\swrong{S_{\rm wrong}}
\def\lwrong{\Scl_{\rm wrong}}
\def\sw{S_\ssw}
\def\lw{\Scl_\ssw}

\def\hf{\frac{1}{2}}
\def\nth#1{\frac{1}{#1}}
\def\Avg#1{\langle #1 \rangle}
\def\Box{{\vbox {\hrule height 0.6pt\hbox{\vrule width 0.6pt\hskip 3pt
        \vbox{\vskip 6pt}\hskip 3pt \vrule width 0.6pt}\hrule height 0.6pt}}}
\def\roughly#1{\mathrel{\raise.3ex\hbox{$#1$\kern-.75em\lower1ex\hbox{$\sim$}}}}
\def\lsim{\roughly<}
\def\gsim{\roughly>}
\def\Tr{\hbox{Tr}\,}
\def\Dslsh{\hbox{/\kern-.6700em\it D}} 
\def\plslsh{\hbox{/\kern-.5300em$\partial$}}

\def\msbar{{\overline{MS}}}
\def\dsbar{{\overline{DS}}}
\def\fpi{F_\pi}
\def\mpl{M_p}
\def\mw{M_\ssw}

\title{Introduction to Effective Field Theory}

\markboth{Effective Field Theories}{Effective Field Theories}

\author{C.P. Burgess
\affiliation{Department of Physics \& Astronomy, McMaster
University,\\ 1280 Main Street West, Hamilton, Ontario, Canada, L8S 4M1\\
{\it and}\\ Perimeter Institute, 31 Caroline Street North,\\
Waterloo, Ontario, Canada, N2L 2Y5}}


\begin{abstract}
This review summarizes Effective Field Theory techniques, which
are the modern theoretical tools for exploiting the existence of
hierarchies of scale in a physical problem. The general
theoretical framework is described, and explicitly evaluated for a
simple model. Power-counting results are illustrated for a few
cases of practical interest, and several applications to Quantum
Electrodynamics are described.
\end{abstract}

\maketitle

\section{Introduction}

It is a basic fact of life that Nature comes to us in many scales.
Galaxies, planets, aardvarks, molecules, atoms and nuclei are very
different sizes, and are held together with very different binding
energies. Happily enough, it is another fact of life that we don't
need to understand what is going on at all scales at once in order
to figure out how Nature works at a particular scale. Like good
musicians, good physicists know which scales are relevant for
which compositions.

The mathematical framework which we use to describe nature ---
quantum field theory --- itself shares this basic feature of
Nature: it automatically limits the role which smaller distance
scales can play in the description of larger objects. This
property has many practical applications, since a systematic
identification of how scales enter into calculations provides an
important tool for analyzing systems which have two very different
scales, $m \ll M$. In these systems it is usually profitable to
expand quantities in the powers of the small parameter, $m/M$, and
the earlier this is done in a calculation, the more it is
simplified.

This review intends to provide a practical introduction to the
technique of Effective Field Theory, which is the main modern tool
for exploiting the simplifications which arise for systems which
exhibit a large hierarchy of scales
\cite{EFT,EFTTexts,EFTDeg,EFTNR}. The goal is to provide an
overview of the theoretical framework, but with an emphasis on
practical applications and concrete examples. The intended
audience is assumed to be knowledgable in the basic techniques of
quantum field theory, including its path-integral formulation.

Although it is not the main focus, one of the more satisfying
threads which I hope you'll find running through this review is
the picture which emerges of the physics underlying the technique
of renormalization. Renormalization is a practice which used to be
widely regarded as distasteful, and so was largely done in the
privacy of one's own home. That has all changed. As used in
effective field theories renormalizing is not only respectable, it
is often the smart thing to do when extracting the dependence of
physical quantities on large logarithms of scale ratios, $\sim
\log(M/m)$.

Another attractive conceptual spinoff of effective field theory
techniques is the understanding they provide of the physical
interpretation of {\it nonrenormalizable} theories, like
Einstein's General Theory of Relativity. Although much has been
made about the incompatibility of gravity and quantum mechanics,
we shall find the quantization of nonrenormalizable theories can
make perfect sense provided they are only applied to low-energy
predictions.

\subsection{A toy model}

In order make the discussion as concrete as possible, consider a
system involving two spinless particles, $l$ and $H$, with one ---
$l$ --- being very light compared with the other --- $H$. Taking
the classical action for the system to be the most general
renormalizable one consistent with the discrete symmetry $l \to -
l$ gives\footnote{We use units $\hbar = c = k_B = 1$ and adopt the
`mostly plus' metric signature.}
\eq \label{toylagr}
    S_c[l,H] = -\int d^4x \; \left[ \hf \;\Bigl(
    \partial_\mu l \,  \partial^\mu l +  \partial_\mu H \,
    \partial^\mu H \Bigr) + V(l,H) \right],
\eeq
where the interaction potential is
\eq \label{toypot}
    V(l,H) = \hf \, m^2 l^2 + \hf \, M^2 H^2
    + \frac{g_l}{4!} \,l^4 + \frac{g_h}{4!} \, H^4 +
    \frac{g_{l h}}{4} l^2 H^2 + \frac{\tilde{m}}{2} \, l^2 H
    + \frac{\tilde{g}_h M}{3!} \, H^3 .
\eeq
$M$ and $m$ denote the two particle masses, and we imagine the
three dimensionful quantities --- $m$, $\tilde{m}$ and $M$ --- to
satisfy $M \gg m, \tilde{m}$, in order to ensure a large hierarchy
of scales.

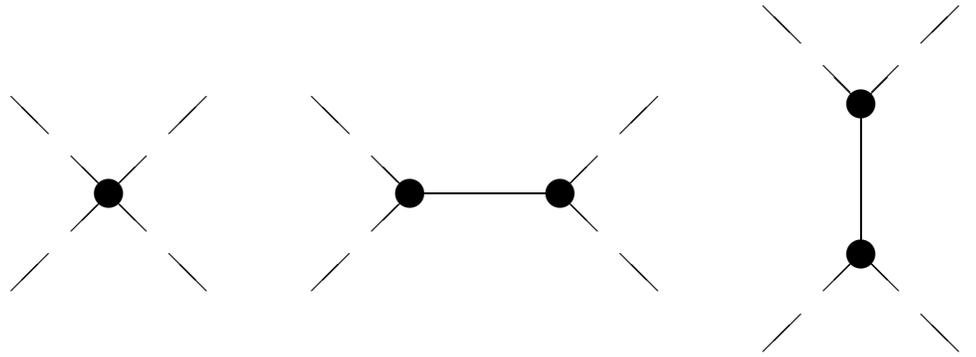
\begin{figure}
\setlength{\unitlength}{1mm} \vspace{2cm}
\centerline{%
\begin{picture}(50,30)
    \put(-20,20){\circle*{4}}
    \multiput(-20,20)(-8,8){2}{\line(-1,1){5}}
    \multiput(-20,20)(-8,-8){2}{\line(-1,-1){5}}
    \multiput(-20,20)(8,8){2}{\line(1,1){5}}
    \multiput(-20,20)(8,-8){2}{\line(1,-1){5}}
    \multiput(20,20)(20,0){2}{\circle*{4}}
    \multiput(20,20)(-8,8){2}{\line(-1,1){5}}
    \multiput(20,20)(-8,-8){2}{\line(-1,-1){5}}
    \multiput(40,20)(8,8){2}{\line(1,1){5}}
    \multiput(40,20)(8,-8){2}{\line(1,-1){5}}
    \put(20,20){\line(1,0){20}}
    \multiput(80,12)(8,-8){2}{\line(1,-1){5}}
    \multiput(80,12)(-8,-8){2}{\line(-1,-1){5}}
    \put(80,12){\circle*{4}}
    \put(80,12){\line(0,1){20}}
    \put(80,32){\circle*{4}}
    \multiput(80,32)(-8,8){2}{\line(-1,1){5}}
    \multiput(80,32)(8,8){2}{\line(1,1){5}}
\end{picture}}
\caption{The Feynman graphs which contribute to two-body
light-particle scattering at tree level in the Toy Model. Solid
(dashed) lines represent the heavy (light) scalar.}
\label{2bdyFgraphsfig}
\end{figure}

Now imagine computing a low-energy physical process in this model,
which we take for simplicity to be two-body $l-l$ scattering at
centre-of-mass energies much smaller than the heavy-particle mass:
$E_{\rm cm} \ll M$. The Feynman graphs which give rise to this
scattering at tree level are given in Fig.~\pref{2bdyFgraphsfig}.

The $S$-matrix element which follows from these graphs may be
written:
\eq \label{smatrixform}
    S(p_1,p_2;p_3,p_4) = i (2 \pi )^4
    \delta^4 (p_1 + p_2 - p_3 - p_4) \; \Sca (p_1,p_2,p_3,p_4),
\eeq
where
\eqa \label{Aform}
    \Sca (p_1,p_2,p_3,p_4) &=& - g_l +
    \tilde{m}^2 \left[ {1 \over (p_1 -  p_3)^2 + M^2} + {1 \over (p_1
    -  p_4)^2 + M^2} \right. \nn\\
    && \qquad\qquad\qquad\qquad\qquad \left. +
    {1 \over (p_1 +  p_2)^2 + M^2} \right] \nn\\
    &\approx& - g_l + {3 \,\tilde{m}^2 \over M^2} +
    {4 \, \tilde{m}^2 m^2 \over M^4} + \Sco\left(M^{-6}\right) \; .
\eeqa
This last (approximate) equality assumes momenta and energies to
be much smaller than $M$. The final line also uses the identity
$(p_1 - p_4)^2 + (p_1 - p_3)^2 + (p_1 + p_2)^2 = - 4 m^2$, which
follows from the mass-shell condition $p^2 = - m^2$.

There are two points to be emphasized about this last expression.
First, this scattering amplitude (and all others) simplifies
considerably in the approximation that powers of $m/M$,
$\tilde{m}/M$ and $E_{\rm cm}/M$ may be neglected. Second, the
result up to $O(M^{-4})$ could be obtained for a theory involving
only $l$ particles interacting through the following `effective'
potential:
\eq \label{eqpot}
    V^{(4)}_{\rm eff} = \frac{1}{4!}\left( g_l -  {3 \, \tilde{m}^2
    \over M^2} - {4 \,\tilde{m}^2 m^2 \over M^4} \right) \; l^4.
\eeq
Even more interesting, the $O(M^{-4})$ contribution to {\it all}
other observables involving low-energy $l$ scattering are
completely captured if the following terms are added to the above
potential:
\eq \label{eqpot1}
    V^{(6)}_{\rm eff} = \frac{\tilde{m}^2}{M^4} \left( \frac{g_{l h}}{16}
    - \frac{g_l}{6} \right) \; l^6.
\eeq

There are several reasons why it is very useful to establish at
the outset that the low-energy interactions amongst light
particles are described, in the large-$M$ limit, by some sort of
effective interactions like eqs.~(\ref{eqpot}) and (\ref{eqpot1}).
Most prosaically, it is obviously much easier to calculate more
complicated processes starting from eqs.~(\ref{eqpot}) and
(\ref{eqpot1}) than by computing the full result using
eqs.~\pref{toylagr} and \pref{toypot}, and only {\it then}
expanding in powers of $1/M$. For more difficult calculations ---
such as the cross section for the reaction $6 \, l \to 12 \, l$,
for instance --- such ease of calculation can be the difference
which decides whether a computation is feasible or not.
Furthermore, knowledge of the types of effective interactions
which are possible at a given order in $1/M$ might guide us to
identify which low-energy observables are most (or least)
sensitive to the properties of the heavy particles.

Fortunately, it appears to be a basic property of quantum field
theories that so long as a large hierarchy of masses exists, a
low-energy description in terms of a collection of effective
interactions is indeed possible. The point of this review is to
sketch why this is so, what may be said about its properties, and
how to compute it (if possible) from a full theory of both light
and heavy particles.

\section{General formulation}

A great variety of observables could be used as the vehicle for
illustrating effective field theory techniques, but since these
can be computed from knowledge of the various correlation
functions of the theory, it is convenient to phrase the discussion
in terms of the generating functional for these. In particular, we
focus attention on the generator, $\G$, of {\sl
one-particle-irreducible} (1PI) correlation functions. Although
this quantity is often called the `effective action,' particularly
in older references, we reserve this name for another quantity of
more direct interest which we come to later.

\subsection{The 1PI and 1LPI actions}

We start by reviewing the standard definition for the generating
functional. Consider a theory whose fields are generically denoted
by $\phi$. Our interest in this theory is in the correlation
functions of these fields, since other physical quantities can
generically be constructed from these. These correlations may be
obtained by studying the response of the theory to the application
of an external field, $J(x)$, which couples to $\phi(x)$.

For instance, a path-integral definition of the correlation
function would be
\eq \label{eqthree}
    \Avg{\phi(x_1)\dots\phi(x_k)}_\ssj \equiv e^{-iW[J]} \; \int
    \Scd\phi \; [\phi(x_1) \dots \phi(x_k)] \;
    \exp\left\{ i \int d^4x \; \Bigl[
    \Scl + J \phi \Bigr] \right\},
\eeq
where $\Scl$ denotes the lagrangian density which describes the
system's dynamics, and the quantity $W[J]$ is defined by:
\eq \label{eqone}
    \exp \Bigl\{i W[J] \Bigr\} = \int \Scd\phi \;
    \exp\left\{ i \int d^4x \; \Bigl[ \Scl[\phi] + J \phi \Bigr]
    \right\}.
\eeq
$W[J]$ generates the {\it connected} correlations of the operator
$\phi$, in the sense that
\eq \label{ccorrs}
    \Avg{ \phi(x_1) \cdots \phi(x_k)}_{c,\ssj}
    = (-i)^k {\delta^k W \over \delta J(x_1) \cdots \delta
    J(x_k) }.
\eeq
This can be taken to define the connected part, but it also agrees
with the usual graphical sense of connectedness. When this average
is evaluated at $J=0$ it coincides with the covariant time-ordered
--- more properly, $T^*$-ordered --- vacuum expectation value of $\phi$.

\subsubsection{The 1PI generator:}

One-particle reducible graphs are defined as those connected
graphs which can be broken into two disconnected parts simply by
cutting a single internal line. One-particle {\it irreducible}
(1PI) graphs are those connected graphs which are not one-particle
reducible.

A non-graphical formulation of 1-particle reducibility of this
sort can be had by performing a Legendre transformation on the
functional $W[J]$ \cite{1PI}. With this choice, if the mean field,
$\varphi$, is defined by
\eq \label{eqtwo}
    \varphi(J) \equiv { \delta W \over \delta J} =
    \Avg{\phi(x)}_\ssj,
\eeq
then the Legendre transform of $W[J]$ is defined to be the
functional $\Gamma[\varphi]$, where
\eq \label{eqfour}
    \Gamma[\varphi] \equiv W[J(\varphi)]
    - \int d^4x \; \varphi \, J. \eeq
Here we imagine $J(\varphi)$ to be the external current that is
required to obtain the expectation value $\Avg{\phi}_\ssj =
\varphi$, and which may be found, in principle, by
inverting\footnote{For simplicity, the inversion of
eq.~(\ref{eqtwo}) is assumed to be possible.} eq.~(\ref{eqtwo}).
If $\Gamma[\varphi]$ is known, $J(\varphi)$ can be found by
directly differentiating the defining equation for
$\Gamma[\varphi]$, which gives:
\eq \label{eqfive}
    {\delta \Gamma[\varphi] \over \delta \varphi} + J = 0.
\eeq

This last equation implies another important property for
$\G[\varphi]$: its stationary point specifies the expectation
value of the original operator, $\Avg{\phi(x)}_{J = 0}$. This can
be seen from eqs.~(\ref{eqtwo}) and (\ref{eqfive}) above. For
time-independent field configurations, $u$, this argument can be
sharpened to show that $\Gamma[\varphi]$ is the minimum
expectation value of the system's Hamiltonian, given that the
expectation of the field $\phi(x)$ is constrained to equal
$\varphi(x)$ \cite{EAenergy}.

A graphical representation for $\Gamma[\varphi]$ can be obtained
by setting up a path-integral representation for it. An expression
for $\Gamma[\varphi]$ as a path integral is found by combining the
definitions of eqs.~(\ref{eqone}) and (\ref{eqfour}):
\eqa \label{eqsix}
    \exp \Bigl\{i \Gamma[\varphi] \Bigr\} &=& \int \Scd\phi
    \; \exp\left\{ i \int d^4x \; \Bigl[ \Scl(\phi) + J \left(
    \phi - \varphi \right) \Bigr] \right\} \nn\\
    &=& \int \Scd \hat\phi
    \; \exp\left\{ i \int d^4x \; \Bigl[ \Scl(\varphi + \hat\phi) + J
    \hat\phi \Bigr] \right\} .
\eeqa

At first sight this last equation is a doubtful starting point for
calculations since it gives only an implicit expression for
$\G[\varphi]$. Eq.~\pref{eqsix} is only implicit because the
current, $J$, appearing on its right-hand side must itself be
given as a function of $\varphi$ using eq.~(\ref{eqfive}): $J = -
(\delta \Gamma / \delta \varphi)$, which again depends on
$\Gamma[\varphi]$. The implicit nature of eq.~(\ref{eqsix}) turns
out not to be an obstacle for computing $\Gamma[u]$, however. To
see this recall that a saddle-point evaluation of eq.~\pref{eqone}
gives $W[J]$ as the sum of all connected graphs that are
constructed using vertices and propagators built from the
classical lagrangian, $\Scl$, and having the currents, $J$, as
external lines. But $\Gamma[\varphi]$ just differs from $W[J]$ by
subtracting $\int d^4x \, J \varphi$, and evaluating the result at
the specific configuration $J[\varphi] = - (\delta \Gamma / \delta
\varphi)$. This merely lops off all of the 1-particle reducible
graphs, ensuring that $\Gamma[\varphi]$ is given by summing
1-particle irreducible graphs.

This gives the following result for $\G[\varphi]$ semiclassically:
\eq \label{clcalg}
    \Gamma[\varphi] =  S[\varphi] + \Gamma_1[\varphi]
    + \hbox{(Feynman Graphs)}.
\eeq
Here, the leading (tree-level) term is the classical action: $S =
\int \Scl (\varphi) \; d^4x$. The next-to-leading result is the
usual one-loop functional determinant,
\eq
    \Gamma_1 = \mp \; \hf \; \log\det\left[ \delta^2
    S/\delta\phi(x) \delta\phi(y) \right]_{\phi = \varphi},
\eeq
of the quadratic part of the action expanded about the
configuration $\phi = \varphi$. The sign, $\mp$, is $-$ for
bosonic fields and $+$ for fermionic ones.

Finally, the `Feynman Graphs' contribution denotes the sum of all
possible graphs which ($i$) involve two or more loops; ($ii$) have
no external lines (with $\varphi$-dependent internal lines
constructed by inverting $\delta^2 S/\delta\phi(x) \delta\phi(y)$
evaluated at $\phi = \varphi$); and ($iii$) are one-particle
irreducible (1PI).

\subsubsection{Light-particle correlations:}

We now specialize this general framework to the specific model
described above containing the two scalar fields, working in the
limit where one is much more massive than the other. If we couple
an external current, $j$, to $l$ and another, $J$, to $H$, then
the generator of 1PI correlations in this model, $\G[\ell,h]$,
depends on two scalar variables, $\ell = \Avg{l}_{jJ}$ and $h =
\Avg{H}_{jJ}$. The correlation functions it generates can be used
to construct general scattering amplitudes for both $l$ and $H$
particles by using standard techniques.

Our interest is in determining the dependence on the heavy mass,
$M$ of low-energy observables, where `low energy' here means those
observables for  which all of the particles involved have
centre-of-mass (CM) momenta and energies which satisfy $p,E \ll
M$. If only $l$ particles are initially present in any scattering
process at such low energies, then no heavy particles can ever
appear in the final state since there is not enough energy
available for their production. As a result, the heavy-field
correlations can be ignored, and it suffices to consider only the
generator of 1PI correlations exclusively for the light fields in
the problem.

Since $H$-correlators are of no interest to us we are free to set
$J = 0$ when evaluating $\Gamma[\ell,h]$, since we never need
differentiate with respect to $J$. As the previous sections show,
the condition $J=0$ is equivalent to evaluating $\Gamma[\ell,h]$
with its argument, $h$, evaluated at the stationary point, $h =
\hbr(\ell)$, where $\delta \Gamma/\delta h$ vanishes.

Furthermore, if only low-energy observables are of interest we may
also restrict the external current $j$ to be slowly varying in
space, so that its Fourier transform only has support on
low-momentum modes for which $p,E \ll M$. As above for $J$ and
$h$, the vanishing of these modes of $j$ correspond to using the
condition $\delta \Gamma/\delta \ell = 0$ to eliminate the
high-frequency components of $\ell$ in terms of the low-frequency
components. The restriction of $\Gamma[\ell,h]$ with these two
conditions, denoted $\gamma[\ell]$, in principle contains all of
the information required to compute any low-energy observable.

How is $\gamma[\ell]$ computed? Recall that $\G[\ell,h]$, obtained
by performing a Legendre transformation on both $j$ and $J$, is
obtained by summing over all 1PI graphs in the full theory. But
$\gamma[\ell]$ differs from $\Gamma[\ell,h]$ only by setting both
$J$ and the short-wavelength components of $j$ to zero (rather
than to the configurations $J = - \delta \Gamma/\delta h$ and $j =
- \delta \Gamma/\delta \ell$). Since it is the currents which are
responsible for cancelling out the one-particle reducible graphs
in $\Gamma[\ell,h]$, we see that setting currents to vanish simply
means that this cancellation does not take place.

We see from this that $\gamma[\ell]$ is given by the sum of
one-{\it light}-particle-irreducible (1LPI) graphs: \ie\ the
graphs which contribute to $\gamma[\ell]$ are one-particle
irreducible only with respect to cutting low-momentum,
low-frequency light-particle, $l$, lines, but are one-particle
reducible with respect to cutting high-momentum $l$ lines or $H$
lines having any momentum.

To make these manipulations concrete, the remainder of this
section derives explicit expressions for both $\G[\ell,h]$ and
$\gamma[\ell]$ in the toy model, working to the tree and one-loop
approximations.

\subsubsection{Tree-level calculation:}

Suppose both $\G[\ell,h]$ and $\gamma[\ell]$ are computed
approximately within a semiclassical loop expansion: $\G = \G^t +
\G_{\oneloop} + \cdots$ and $\gamma = \gamma^t + \gamma^\oneloop +
\cdots$. Explicit formulae for both are particularly simple at
tree level since the tree approximation, $\G^t$, to $\G$ is simply
given by evaluating the classical action at $l = \ell$ and $H =
h$: $\G^t[\ell,h] = S[\ell,h]$. The tree approximation for
$\gamma^t[\ell]$ is therefore obtained by solving the classical
equations of motion for $h$ --- and the high-frequency part of
$\ell$) as a function of (the low-frequency part of) $\ell$ ---
and then substituting this result, $\hbr_t(\ell)$, back into the
classical action.

We next compute the resulting expression for $\gamma^t[\ell]$ in
the limit that $M$ is much larger than all of the other scales of
interest. The classical equation of motion for $h$ which is
obtained from eqs.~(\ref{toylagr}) and (\ref{toypot}) is
\eq \label{cleqforh}
    \Box\, h - M^2 h - \hf \, g_{l h} \ell^2 h
    - \frac{g_h}{3!} \, h^3 - \frac{\tilde{m}}{2} \, \ell^2 -
    \frac{\tilde{g}_h M}{2} \, h^2 = 0,
\eeq
so the solution, $\hbr_t(\ell)$ can be formally written:
\eq \label{hsoln}
    \hbr_t(\ell) = \left( \Box - M^2 -
    \frac{g_{l h}}{2} \,  \ell^2 \right)^{-1} \; \left(
    \frac{\tilde{m}}{2} \, \ell^2 + \frac{g_h}{3!} \, \hbr^3 +
    \frac{\tilde{g}_h M}{2} \, \hbr^2 \right) .
\eeq
This may be solved perturbatively in powers of $g_h$, $\tilde g_h$
and $1/M$, with the leading contribution obtained by taking $\hbr
= 0$ on the right-hand side. This leads to the explicit solution
\eqa \label{hsoln}
    \hbr_t(\ell) &=& \left[ - \, \nth{M^2}
    - \nth{M^4} \left( \Box - \frac{g_{l
    h}}{2} \, \ell^2  \right)  + \cdots \right]
    \left(\frac{\tilde{m}}{2} \, \ell^2
    \right)  + \Sco(g_h, \tilde{g}_h) \nn \\
    &=& - \frac{\tilde{m}}{2M^2} \, \ell^2  + \frac{g_{l h}
    \tilde{m}}{4 M^4} \, \ell^4 - \frac{\tilde{m}}{2M^4} \Box \left(
    \ell^2 \right) + \Sco\left( {1 \over M^5} \right) ,
\eeqa
where the leading contributions involving nonzero $g_h$ and
$\tilde g_h$ contribute at $\Sco(M^{-5})$.

Substituting this last result back into the classical action gives
the tree level expression for the generating functional,
$\gamma[\ell]$, as an expansion in powers of $1/M$. We find the
result $\gamma^t[\ell] = \gamma^t_{0}[\ell] + \gamma^t_2[\ell]/M^2
+ \gamma^t_4[\ell]/M^4 + \cdots$, with:
\eqa \label{wefind}
    \gamma^t_0[\ell] &=& \int d^4x \; \left( -
    \hf \; \partial_\mu \ell \,
    \partial^\mu \ell - \frac{m^2}{2} \, \ell^2 - \frac{g_l}{4!} \,
    \ell^4 \right),\nn\\
    {\gamma^t_2[\ell] \over M^2} &=& \int d^4x \; \left(
    \frac{\tilde{m}^2}{8M^2} \;
    \ell^4 \right), \\
    {\gamma^t_4[\ell] \over M^4} &=& \int d^4x \; \left( - \,
    \frac{\tilde{m}^2}{2M^4} \; \ell^2 \, \partial_\mu \ell \,
    \partial^\mu \ell -  \frac{g_{l h} \, \tilde{m}^2}{16 M^4} \;
    \ell^6 \right) . \nn
\eeqa

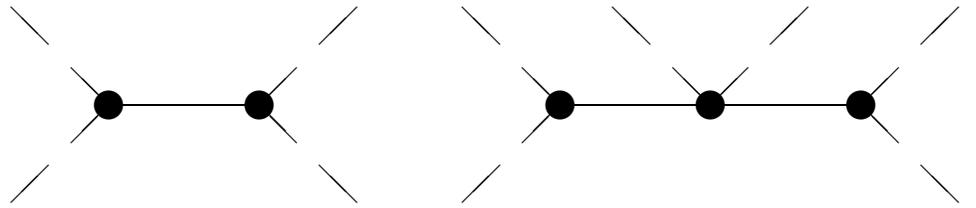
\begin{figure}
\setlength{\unitlength}{1mm}
\centerline{%
\begin{picture}(50,30)
    \multiput(-20,12)(20,0){2}{\circle*{4}}
    \multiput(-20,12)(-8,8){2}{\line(-1,1){5}}
    \multiput(-20,12)(-8,-8){2}{\line(-1,-1){5}}
    \multiput(0,12)(8,8){2}{\line(1,1){5}}
    \multiput(0,12)(8,-8){2}{\line(1,-1){5}}
    \put(-20,12){\line(1,0){20}}
    \multiput(40,12)(40,0){2}{\circle*{4}}
    \multiput(40,12)(-8,8){2}{\line(-1,1){5}}
    \multiput(40,12)(-8,-8){2}{\line(-1,-1){5}}
    \multiput(80,12)(8,8){2}{\line(1,1){5}}
    \multiput(80,12)(8,-8){2}{\line(1,-1){5}}
    \put(40,12){\line(1,0){40}}
    \put(60,12){\circle*{4}}
    \multiput(60,12)(-8,8){2}{\line(-1,1){5}}
    \multiput(60,12)(8,8){2}{\line(1,1){5}}
\end{picture}}
\caption{The Feynman graphs which contribute the corrections to
the tree level generating functional, $\gamma[\ell]$, to order
$1/M^4$. Solid (dashed) lines represent heavy (light) scalars.}
\label{FgraphsM6}
\end{figure}

These expressions for $\gamma^t[\ell]$ also have a simple
representation  in terms of Feynman graphs. All of the
$M$-dependent terms may be obtained by summing connected tree
diagrams having only $\ell$ particles as external lines and only
$h$ particles for internal lines, and then Taylor expanding all of
the $h$ propagators in powers of $1/M$. The graphs of this type
which contribute up to and including $O(1/M^6)$ are given
explicitly in Fig.~\pref{FgraphsM6}.

This calculation brings out several noteworthy points:

\smallskip\noindent$\bullet$ {\it Decoupling:}
All of the $M$ dependence in $\gamma^t[\ell]$ vanishes as $M \to
\infty$, reflecting (at tree level) the general result that
particles decouple from low-energy physics in the limit that their
mass becomes large.

\smallskip\noindent$\bullet$ {\it Truncation:}
The only part of $\gamma^t[\ell]$ which survives as $M \to \infty$
consists of those terms in the classical action which are
independent of the heavy field, $h$. That is, at tree level
$\gamma^t_0[\ell]$ is obtained from the classical action,
$S[\ell,h]$, simply by truncating it to $h=0$: $\gamma^t_0[\ell] =
S[\ell,0]$. Notice that this relies on two assumptions: the
absence of $M^3 H$ terms in the potential, and the condition
$\tilde{m} \ll M$. (Even though $\hbr^t(\ell)$ would still vanish
like $1/M$ if $\tilde{m}$ were $O(M)$, this is not enough to
ensure the vanishing of terms like $M^2 (\hbr^t)^2$. This latter
condition often breaks down in supersymmetric theories
\cite{SUSYtrunc}.)

\smallskip\noindent$\bullet$ {\it Locality:}
Although the exact expression for $\hbr_t(\ell)$, and hence for
$\gamma^t[\ell]$, involves nonlocal quantities such as $G(x,x') =
\langle x | \left(-\Box + M^2\right)^{-1} | x' \rangle$, the
entire result becomes local once these are expanded in powers of
$1/M$ because of the locality of propagators in this limit:
\eq
    G(x,x') = \int \frac{d^4p}{(2\pi)^4} \left[
    \frac{e^{ip(x-x')}}{p^2 + M^2} \right]
    = \left[ \frac{1}{M^2} + \frac{\Box}{M^4} + \cdots \right]
    \, \delta^4(x-x') .
\eeq
This locality is ultimately traceable to the uncertainty
principle. The key observation is that the $M$-dependent
interactions in $\gamma[\ell]$ express the effects of virtual
heavy particles whose energies $E_h \ge M$, are much higher than
those of any of the light particles whose scatterings are under
consideration in the low-energy limit. Indeed, it is precisely the
high energy of these particles which precludes their being
inadvertently produced as real final-state particles in any
low-energy scattering process, and so guarantees that they only
mediate transitions amongst the light-particle states.

But the virtual contribution of heavy particles to low-energy
processes can nevertheless occur within perturbation theory
because the uncertainty principle permits the violation of energy
conservation required for their production, provided it takes
place only over short enough times, $\Delta t \lsim 1/\Delta E_h
\lsim 1/M$. As a consequence, from the low-energy perspective the
influence of heavy particles appears to be instantaneous (\ie\
local in time). The uncertainty principle similarly relates the
momentum required to produce heavy virtual particles with the
distances over which they can travel, thereby making their
influence also {\it local} in space.\footnote{We assume a
relativistic dispersion relation, $E^2 = p^2 + m^2$, so states
having large momentum also have high energy, and must therefore be
excluded from the low-energy theory.}

The influence at low energies, $E$, of very massive particles, $M
\gg E$, can therefore generically be reproduced to a fixed order
in $E/M$ by local interactions involving only the particles which
are present at low energies. It is precisely this locality which
makes possible the construction of a low-energy effective theory
which accurately describes these virtual effects.

\smallskip\noindent$\bullet${\it Redundant interactions:}
At face value eq.~\pref{wefind} is not precisely the same as the
effective potential, $V^{(4)} + V^{(6)}$, encountered in
eqs.~\pref{eqpot} and \pref{eqpot1} of the introduction. This
apparent difference is illusory, however, because the difference
can be removed simply by performing a field redefinition.

To see how this works, suppose we have an action, $S[\phi^i]$,
given as a series in some small parameter $\veps$:
\eq \label{actionseries}
    S[\phi^i] = S_0[\phi^i] + \veps \;
    S_1[\phi^i] + \veps^2 \; S_2[\phi^i] + \cdots \, ,
\eeq
where $\veps$ could be a small loop-counting parameter or a small
energy ratio, $E/M$. Suppose also that somewhere amongst the
interactions which appear in the $O(\veps^n)$ contribution,
$S_n[\phi^i]$, there is a term, $S_n^\ssr[\phi^i]$, which vanishes
when the fields, $\phi^i$ are chosen to satisfy the equations of
motion for the lowest-order action, $S_0[\phi^i]$. In equations:
\eq \label{termvanishes}
    S_n^\ssr[\phi] = \int d^4x \; f^i(x) \,
    {\delta S_0 \over \delta \phi^i(x)},
\eeq
where the coefficients $f^i(x)$ are ultra-local functions of the
fields and their derivatives at the spacetime point $x$.

The claim is that, {\em to order $\veps^n$}, any such interaction,
$S_n^\ssr[\phi^i]$, can be removed by performing a field
redefinition without altering any of the other terms at lower or
equal order, and so can have no physical consequences. The
required field redefinition is:
\eq \label{reqfredef}
    \phi^i(x) \to \tilde\phi^i(x) = \phi^i(x) -
    \veps^n \; f^i(x),
\eeq
since under this redefinition the $O(\veps^n)$ terms in the action
vary into:
\eq \label{changeins}
    S[\phi^i] \to S[\tilde\phi^i] = S[\phi^i] -
    \veps^n \int d^4x \;  f^i(x) \, {\delta S_0 \over \delta
    \phi^i(x)} + O(\veps^{n+1}) .
\eeq
Clearly, to $O(\veps^n)$ the sole effect of this redefinition is
simply to cancel $S_n^\ssr[\phi^i]$.

Applying this reasoning to $\gamma[\ell]$, above, notice that
integrating by parts gives
\eq
    \int d^4x \; \ell^2 \, \partial_\mu \ell \, \partial^\mu \ell
    = - \frac13 \int d^4x \; \ell^3 \, \Box \, \ell  ,
\eeq
and so using the lowest-order equations of motion for $\ell$
derived from $\gamma^t_0[\ell]$ -- \ie\ $\Box \, \ell = m^2 \ell +
g_l \, \ell^3/3!$ -- implies
\eq
    \int d^4x \; \ell^2 \, \partial_\mu \ell \, \partial^\mu \ell
    = - \frac13 \int d^4x \; \left( m^2 \, \ell^4 +
    \frac{g_l}{3!} \, \ell^6 \right) .
\eeq
Using this in eq.~\pref{wefind} reproduces the potential of
eqs.~\pref{eqpot} and \pref{eqpot1}.

\subsubsection{One-loop calculation:}

More can be learned by examining some of the subdominant terms in
the loop expansion for $\gamma[\ell]$.

At face value keeping one-loop corrections changes $\gamma[\ell]$
in two different ways: through the one-loop corrections to the
functional form of $\G[\ell,h]$; and through the corrections that
these imply for the stationary point, $\hbr(\ell)$. In practice,
however, the correction to $\hbr(\ell)$ does not contribute to
$\gamma[\ell]$ at one loop. To see this use the expansions $\G =
\G^t + \G^{\oneloop} + \cdots$ and $\hbr = \hbr^t +
\hbr^{\oneloop} + \cdots$ into the definition $\gamma[\ell]$. To
one-loop order this gives
\eqa \label{expns}
    \gamma[\ell] &=& \G^t[\ell,\hbr_t + \hbr^{\oneloop}] +
    \G^{\oneloop}[\ell,\hbr^t] + \cdots \nn\\
    &=& \left[ \G^t[\ell,\hbr^t] + \int d^4x \;
    \left( {\delta \G^t \over \delta h} \right)_{\hbr^t}
    \hbr^\oneloop \right] + \G^{\oneloop}[\ell,\hbr^t]  + \cdots \nn\\
    &=& \gamma^t[\ell] + \G^{\oneloop}[\ell,\hbr^t]  + \cdots ,
\eeqa
where the second equality follows by expanding $\G^t$ about $h =
\hbr^t$, keeping only terms of tree and one-loop order, and the
last equality uses the fact that $\delta \G^t / \delta h = 0$ when
evaluated at $\hbr^t$, together with the tree-level result,
$\gamma^t[\ell] = \G^t[\ell,\hbr^t(\ell)]$. Eq.~\pref{expns}
states that the one-loop approximation to $\gamma$ is obtained by
evaluating $\G^\oneloop$ at the {\em tree-level} configuration
$\hbr^t$: \ie\ $\gamma^{\oneloop}[\ell] =
\G^{\oneloop}[\ell,\hbr^t(\ell)]$.

To proceed we require the one-loop contribution, $\G_{\oneloop}$,
which is given by:
\eq \label{oneloopexpression}
    \G_{\oneloop}[\ell,h] = \frac{i}{2}
    \log \det \left[ \matrix{ (- \Box + V_{\ell\ell})/\mu^2 & V_{\ell
    h}/\mu^2 \cr V_{\ell h}/\mu^2 & (- \Box + V_{hh})/\mu^2 \cr}
    \right],
\eeq
where $\mu$ is an arbitrary scale required on dimensional grounds,
and the matrix of second derivatives of the scalar potential is
given by
\eqa \label{scndderivs}
    \Bfv &=& \pmatrix{V_{\ell\ell} & V_{\ell h} \cr V_{\ell h}
    & V_{hh} \cr} \nn\\
    &=& \pmatrix{ m^2 + \frac{g_l}{2} \, \ell^2 + \frac{g_{l
    h}}{2} \, h^2 + \tilde m h & g_{l h} \ell h + \tilde{m} \ell
    \cr g_{l h} \ell h +
    \tilde m \ell & M^2 + \tilde{g}_h \, M h + \frac{g_h}{2} \, h^2 +
    \frac{g_{l h}}{2} \, \ell^2 \cr}.
\eeqa

Evaluating the functional determinant in the usual way gives an
expression which diverges in the ultraviolet. For the two-scalar
theory under consideration these divergences only appear (at one
loop) in that part of $\G[\ell,h]$ which does not depend on
derivatives of $\ell$ or $h$ (\ie\ the scalar `effective
potential'). If the divergent terms are written $\G_{\rm div} = -
\int d^4x \; V_{\rm div}$, then $V_{\rm div}$ is:
\eq \label{divterms}
    V_{\rm div} = {1 \over 32 \pi^2} \left[ C +
    (V_{\ell\ell} + V_{hh}) \Lambda^2
    - \hf \, (V^2_{\ell\ell} + V^2_{hh}
    + 2 V^2_{\ell h}) \; L \right]
\eeq
where $C$ is a field-independent, divergent constant, and $L =
\log\left( \Lambda^2  / \mu^2 \right)$. $\Lambda$ is an
ultraviolet cutoff which has been used to regulate the theory, and
which is assumed to be sufficiently large compared to all other
scales in the problem that all inverse powers of $\Lambda$ can be
neglected.

Alternatively, using dimensional regularization instead of an
ultraviolet cutoff leads to the same expressions as above, with
two changes: ($i$) all of the terms proportional to $C$ or to
$\Lambda^2$ are set to zero, and ($ii$) the logarithmic divergence
is replaced by $L \to 2/(4 - n) = 1/\varepsilon$, where $n = 4 -
2\varepsilon$ is the dimension of spacetime.

All of the divergences can be absorbed into renormalizations of
the parameters of the lagrangian, by defining the following
renormalized couplings:
\eqa \label{RG}
    A_\ssr &=& A_0 + {1 \over 32 \pi^2} \; \left[C +
    (M^2 + m^2) \Lambda^2  - \hf \,
    (M^4 + m^4) \; L \right], \nn\\
    B_\ssr &=& B_0 + {1 \over 32 \pi^2}\;  \left[ (\tilde{m} +
    \tilde{g}_h M) \Lambda^2 -
    (m^2 \tilde{m} + \tilde{g}_h M^3) \; L \right] , \nn\\
    m^2_\ssr &=& m^2 + {1 \over 32 \pi^2}\;  \left[ (g_l + g_{l
    h}) \Lambda^2 -
    (g_l m^2 + g_{l h} M^2 + 2 \tilde{m}^2) \; L  \right] , \\
    M^2_\ssr &=& M^2 + {1 \over 32 \pi^2}\;  \left[ (g_h + g_{l h})
    \Lambda^2 - \Bigl(g_{l h} m^2 + (g_h
    + \tilde{g}_h^2) M^2 + \tilde{m}^2 \Bigr) \; L \right] , \nn\\
    \tilde{m}_\ssr &=& \tilde{m} - \, {1 \over 32 \pi^2} \; (g_l
    \tilde{m} + g_{l h}
    \tilde{g}_h M + 4 g_{l h} \tilde{m}) \; L , \nn\\
    (\tilde{g}_h)_\ssr &=& \tilde{g}_h - \, {3 \over 32 \pi^2} \;
    \left(g_{l h} {\tilde{m}
    \over M} +  g_h \tilde{g}_h \right) \; L  \nn\\
    (g_l)_\ssr &=& g_l - \, {3 \over 32 \pi^2} \; (g_l^2 +
    g_{l
    h}^2) \; L \nn\\
    (g_h)_\ssr &=& g_h  - \, {3 \over 32 \pi^2} \;  (g_h^2 + g_{l
    h}^2) \; L \nn\\
    (g_{l h})_\ssr &=& g_{l h} - \, {1 \over 32 \pi^2} \;
    (g_l + g_h +  4 g_{l h}) g_{l h} \; L . \nn
\eeqa

In these expressions $A_0$ and $B_0$ are the coefficients of the
terms $A + B h$ which do not appear in the classical potential,
but which are not forbidden by any symmetries. By not including
such terms in the classical potential we are implicitly choosing
$A_0$ and $B_0$ to satisfy the renormalization condition $A_\ssr =
B_\ssr = 0$, which can be ensured by appropriately shifting the
renormalized fields. Similarly, the assumed heirarchy of masses
between the light field, $\ell$, and the heavy field, $h$, assumes
the {\em renormalized} quantities satisfy $m_\ssr, \tilde{m}_\ssr
\ll M_\ssr$. Notice that these conditions are `unnatural' in that
they require that large $O(M)$ renormalization corrections must be
cancelled by the bare quantities $A_0$, $B_0$, $m$ and
$\tilde{m}$, in addition to the cancellation of any divergent
parts. In what follows we assume that this renormalization has
been performed, and that all parameters which appear in
expressions are the renormalized quantities --- even though the
subscript `$R$' is not explicitly written.

Once this renormalization has been performed the remaining
expression is finite and its dependence on the heavy mass scale,
$M$, can be identified. As for the tree-level analysis, it is
convenient to organize $\gamma^\oneloop[\ell]$ into an expansion
in powers of $1/M$. A new feature which arises at one loop is that
the dominant term in this expansion now varies as $M^4 \log M$
rather than as $M^0$ when $M \to \infty$. We therefore write:
$\gamma^\oneloop = \gamma^\oneloop_{-4} M^4 + \gamma^\oneloop_{-2}
M^2 + \gamma^\oneloop_{-1} M + \gamma^\oneloop_0 +
\gamma^\oneloop_2 / M^2 + \cdots$.

It turns out that those terms in $\gamma^\oneloop[\ell]$ which
involve derivatives of $\ell$ first appear at order $1/M^2$ in
this expansion.\footnote{This result is a special feature of the
two-scalar system at one loop. At two loops, or at one loop for
other systems, kinetic terms like $\partial_\mu \ell \,
\partial^\mu \ell$ can appear proportional to logarithms of $M$.}
To identify all of the terms which are larger than $O(1/M^2)$ it
therefore suffices to work only with the effective scalar
potential. After performing the renormalizations indicated in
eqs.~\pref{RG}, the one-loop result for the scalar potential
becomes:
\eq \label{effpotexp}
    \gamma_{\rm pot}^\oneloop[\ell] = - \,
    \nth{64 \pi^2} \int d^4x \left[ V_+^2 \log\left( { V_+ \over
    \mu^2} \right) + V_-^2 \log\left({ V_- \over \mu^2} \right)
    \right],
\eeq
where $V_\pm$ denotes the two eigenvalues of the matrix $\Bfv$,
which are given explicitly by:
\eq \label{evalexpr}
    V_\pm = \hf \, \left[ (V_{\ell \ell} +
    V_{hh}) \pm \sqrt{ (V_{\ell \ell} - V_{hh})^2 + 4 V_{\ell h}^2 }
    \right].
\eeq
Our interest is in a potential for which $V_{hh} \gg V_{\ell
\ell},  V_{\ell h}$,  and so these expressions can be simplified
to:
\eqa \label{approxvpm}
    V_+ &\approx& V_{hh} + {V_{\ell h}^2 \over
    V_{hh}} + {V_{\ell \ell}  V_{\ell
    h}^2 \over V_{hh}^2} + O\left({1\over V_{hh}^3} \right), \nn\\
    V_- &\approx& V_{\ell \ell} - {V_{\ell h}^2 \over V_{hh}} -
    {V_{\ell  \ell} V_{\ell h}^2 \over V_{hh}^2} + O\left({1\over
    V_{hh}^3} \right),
\eeqa
and:
\eqa \label{approxeffpotexpr}
    \gamma_{\rm pot}^\oneloop[\ell] &=&
    - \, \nth{64 \pi^2} \int d^4x \left\{ V_{hh}^2 \; \log\left(
    {V_{hh} \over \mu^2} \right) + V_{\ell \ell}^2 \;
    \log\left( {V_{\ell \ell} \over \mu^2} \right) \right. \\
    && \qquad \left. + V_{\ell h}^2 \left[ 1 + 2 \log\left(  {V_{hh}
    \over \mu^2} \right) \right] + {2 V_{\ell \ell} V_{\ell h}^2 \over
    V_{hh}} \; \log\left( {V_{hh} \over V_{\ell \ell}} \right) +
    O\left( {1 \over V_{hh}^2} \right) \right\}. \nn
\eeqa

Using the explicit expressions given above for the scalar
potential, and evaluating the result at $h = \hbr^t(\ell)$, the
first few terms of the $1/M$ expansion at one loop are given by:
\eqa \label{oneloopfinpart}
    M^4 \gamma^\oneloop_{-4} &=& -  \,
    {M^4 \over 64 \pi^2} \;
    \log\left( {M^2 \over \mu^2} \right)  \int d^4x , \nn\\
    M^2 \gamma^\oneloop_{-2}[\ell] &=& - \, { M^2 \over 64 \pi^2}
    \; \left[ \log\left( {M^2 \over \mu^2} \right) + \hf
    \right]  \int d^4x \; g_{l h} \;  \ell^2, \\
    M \gamma^\oneloop_{-1}[\ell] &=& + \, {M \over 64 \pi^2} \; \left[
    \log\left( {M^2 \over \mu^2} \right) + \hf \right]
    \int d^4x \; \tilde{g}_h \tilde{m} \;  \ell^2, \nn\\
    \gamma^\oneloop_{0} &=& -  {1 \over 64 \pi^2} \int
    d^4x \; \left[ \left( \frac{g^2_{l h}}{4} \; \ell^4 + 2
    \tilde{m}^2 \ell^2 \right)
    \log\left(  {M^2 \over \mu^2} \right) \right. \nn\\
    && \qquad \left. + \, {3 g^2_{l h} \over 8} \; \ell^4 +
    \tilde{m}^2 \ell^2  + \left( m^2 +  \frac{g_l^2}{2} \ell^2
    \right)^2 \log\left( {m^2 + \frac{g_l^2}{2} \; \ell^2  \over
    \mu^2 } \right) \right]  . \nn
\eeqa

We see that this shares two of the crucial properties of the
tree-level result:
\noindent$\bullet${\it Decoupling:}
Notice that --- superficially --- the effects of the heavy
particle no longer appear to vanish as $M \to \infty$. However,
all of the terms which grow as $M$ grows have the same form as
does the classical lagrangian, and so they can all be absorbed
into finite renormalizations of $A_\ssr$, $m^2_\ssr$ and
$(g_l)_\ssr$. That is, if we define the new quantities:
\eqa \label{RGprime}
    A_\ssr' &=& A_\ssr + {M^4 \over 64 \pi^2} \;
    \log\left( {M^2 \over \mu^2}
    \right) \nn\\
    (m^2_\ssr)' &=& m^2_\ssr + {1 \over 64 \pi^2} \left\{  M^2 \left(
    g_{l h} -
    \tilde{g}_h {\tilde{m} \over M} \right) \left[ 1 +  2 \log\left( {M^2 \over \mu^2}
    \right)  \right] + 4 \tilde{m}^2 \log\left( {M^2 \over \mu^2}\right) \right\} \nn\\
    (g_l)_\ssr' &=& (g_l)_\ssr + {3 \,g^2_{l h}\over 32 \pi^2} \;
    \log\left( {M^2 \over \mu^2}\right) ,
\eeqa
then the one-loop contribution, $\gamma^\oneloop[\ell]$, to the
1LPI generator becomes:
\eqa \label{oneloopaftrenprime}
    \gamma^\oneloop[\ell] &=& -
    {1 \over 64 \pi^2}  \int d^4x \;
    \left[ {3 g^2_{l h} \over 8} \; \ell^4 + \tilde{m}^2 \ell^2 \right. \\
    && \qquad \qquad \left. + \left( m^2 +  \frac{g_l^2}{2}
    \ell^2 \right)^2 \log\left( {m^2 + \hf \, g_l^2 \, \ell^2 \over
    \mu^2 } \right) \right] + O\left( {1 \over M} \right), \nn
\eeqa
where $m^2$ represents $(m^2_\ssr)'$, and so on.

Clearly, after such renormalizations are performed, all of the
remaining $M$ dependence vanishes in the limit $M \to \infty$.
Provided the values of renormalized couplings are in any case
inferred from experiment, all of the {\em physical} effects of the
heavy particle really are suppressed for large $M$, ensuring the
heavy particle does decouple from physical observables.

\smallskip\noindent$\bullet${\it Locality:}
Since the one-loop action, $\gamma^{\oneloop}[\ell]$, is the
integral over spacetime of a quantity which is evaluated at a
single spacetime point when expanded in inverse powers of $M$, it
shares the locality of the classical result. The underlying source
of this locality is again the uncertainty principle, which
precludes violations of energy and momentum conservation over
large distances -- a result which hinges on our keeping only
states that are defined by their low energy.

\subsection{The Wilson action}

We next reorganize the same calculation, with the goal of making
the $M$ dependence of physical results manifest from the outset.
To this end, suppose we start from the path-integral expression
for the 1LPI generating functional, $\gamma[\ell]$, derived as
eq.~\pref{eqsix} above,
\eq \label{effactionexp}
    \exp \Bigl\{i \gamma[\ell] \Bigr\} = \int
    \Scd l \; \Scd H \; \exp\left\{ i \int d^4x \; \Bigl[ \Scl(\ell + l,H)
    + j l \Bigr] \right\} ,
\eeq
with the external current, $j$, regarded as being defined by
$j[\ell] = - \delta \gamma[\ell]/ \delta \ell$. No similar current
is coupled to the heavy field, $H$, in eq.~\pref{effactionexp}
since our attention is restricted to low-energy processes for
which no heavy particles appear in the initial or final states.

\subsubsection{The Wilson action:}

Now imagine schematically dividing the functional integral into
its low-energy and high-energy parts, $\Scd l \; \Scd H = [\Scd
l]_\lowe [\Scd l \; \Scd H]_\hie$, relative to some arbitrary
intermediate scale, $\lambda$. (For instance, this might be done
by requiring high-energy modes to satisfy $p^2 + m^2
> \lambda^2$ in Euclidean signature.) Using this distinction
between low- and high-energy modes it becomes possible to perform
the functional integration over --- or to `integrate out' --- the
high-energy modes once and for all:
\eq \label{gammafromsw}
    \exp \Bigl\{ i \gamma[\ell] \Bigr\} = \int
    [\Scd l]_\lowe \exp \left\{ i \int d^4x \left[ \lw
    \left( \ell + l_\lowe \right) + j l_\lowe \right]
    \right\} ,
\eeq
where $\sw = \int d^4x \; \lw$ is called the {\em Wilson action},
and is defined as the result of performing the high-energy  part
of the functional integral:
\eq \label{wilsondef}
    \exp \Bigl\{ i \sw \left[
    \ell + l_\lowe \right] \Bigr\} = \int [\Scd
    l \; \Scd H]_\hie \; \exp \Bigl\{ i S \left[\ell + l_\lowe,
    l_\hie, H_\hie \right] \Bigr\} .
\eeq
Eqs.~\pref{gammafromsw} and \pref{wilsondef} are the central
definitions from which the calculation of $\gamma[\ell]$ {\it \`a
la} Wilson proceed.

There are two points about these last expressions which bear
special emphasis. Notice first that $\lw$ appears in
eq.~\pref{gammafromsw} in precisely the same way as would the
classical lagrangian in a theory for which no heavy field existed.
Consequently, once $\lw$ is known it may be used to compute
$\gamma[\ell]$ in the usual way: one must sum over all 1LPI vacuum
Feynman graphs using the interactions and propagators for the
light fields which are dictated by the effective lagrangian
density, $\lw$.

The second point --- which is what makes eq.~\pref{gammafromsw} so
useful --- is that because $\lw$ is computed by integrating only
over high-energy modes, the uncertainty principle guarantees that
it is {\it local} once it is expanded in inverse powers of the
heavy scales. Consequently, to the extent that we work only to a
fixed order in this expansion, we need not worry that
eq.~\pref{wilsondef} will generate arbitrary non-local
interactions.

\subsubsection{The physics of renormalization:}

Eqs.~\pref{gammafromsw} and \pref{wilsondef} share another
beautiful feature. Although the Wilson action depends explicitly
on the scale $\lambda$ which is used in its definition, this
dependence {\it always} drops out of any physical observables. It
must drop out since it only arises from our choice to perform the
calculation in two steps: first integrating modes heavier than
$\lambda$, and then integrating the lighter modes.

In detail, this cancellation arises because $\lambda$ enters into
the low-energy part of the calculation in two ways. The first way
is through the explicit $\lambda$ dependence of all of the
couplings of the Wilson action, $\sw$. However, $\lambda$ also
enters because all of the contributions of virtual particles in
the low-energy theory have their momenta cutoff at the scale
$\lambda$. The $\lambda$ dependence of the couplings in $\sw$ is
just what is required to cancel the $\lambda$'s which enter
through the cutoff.

This entire discussion of $\lambda$-cancellation induces a strong
sense of {\em d\'eja vu}, because it exactly parallels the
traditional renormalization programme wherein the regularization
dependence of divergent loop integrals are cancelled by
introducing regularization-dependent interactions (or
counter-terms) into the classical action, $S$. This similarity
makes it irresistible to regard the original classical action,
$S[l,H]$, as itself being the Wilson action for a yet more
fundamental theory which applies at still higher energies, above
the cutoff $\Lambda$. Any such Wilson action would be used to
compute physical observables in precisely the same way as one
traditionally uses the classical action, including the
renormalization of all couplings to cancel the cutoff dependence
of all observable quantities. The great benefit of adopting this
point of view is the insight it gives into the physical nature of
this cancellation.

\subsubsection{The dimensionally regularized Wilson action:}
\label{dimregWilson}

The Wilson action defined with an explicit cutoff is somewhat
cumbersome for practical calculations, for a variety of reasons.
Cutoffs make it difficult to keep the gauge symmetries of a
problem manifest when there are spin-one gauge bosons (like
photons) in the problem. Cutoffs also complicate our goal of
following how the heavy scale, $M$, appears in physically
interesting quantities like $\gamma[\ell]$, because they muddy the
dimensional arguments used to identify which interactions in $\sw$
contribute to observables order-by-order in $1/M$.

It is much more convenient to use dimensional regularization, even
though dimensional regularization seems to run counter to the
entire spirit of a low-energy action by keeping momenta which are
arbitrarily high. This is not a problem in practice, however,
because the error we make by keeping such high-momentum modes can
itself always be absorbed into an appropriate renormalization of
the effective couplings. This is always possible precisely because
our `mistake' is to keep high-energy modes, whose contributions at
low energies can always be represented using local effective
interactions. Whatever damage we do by using dimensional
regularization to define the low-energy effective action can
always be undone by appropriately renormalizing our effective
couplings.

We are led to the following prescription for defining a
dimensionally-regularized effective action in the two-scalar toy
model. First dimensionally regulate the full theory involving both
fields $l$ and $H$, using for convenience the mass-independent
$\msbar$ renormalization scheme. At one loop this amounts to
renormalizing as in eqs.~\pref{RG}, but with all of the
quartically and quadratically divergent terms set to zero, and
substituting $L \to 1/\varepsilon + k$ in the logarithmically
divergent terms, where $k = \gamma - \log(4\pi)$ and $\gamma =
0.577~215\cdots$ is the Euler-Mascherelli constant.

Next define the effective theory to include only the light field
$l$, also regulated using dimensional regularization. However
rather than using minimal subtraction in the effective theory we
instead renormalize the effective couplings by demanding that they
successfully reproduce the low-energy limit of the full theory,
using for this purpose any convenient set of observables. Once
this {\em matching} calculation has been done, the resulting
effective theory can be used to compute any other quantities as
required. This construction is best understood using the concrete
example of the two-scalar model.

\subsubsection{Tree-Level Calculation:}

Imagine computing both $\gamma$ and $\lw$ within the loop
expansion: $\gamma = \gamma^t + \gamma^\oneloop + \cdots$ and $\lw
= \lw^t + \lw^\oneloop + \cdots$. At tree level the distinction
between  $\sw = \int d^4x \; \lw$, and the 1LPI generator,
$\gamma$, completely degenerates. This is because the tree
approximation to $\gamma[\ell]$ is simply obtained by evaluating
the integrands of eqs.~\pref{gammafromsw} and \pref{wilsondef} at
the classical saddle point. In the present case this implies  that
$\gamma^t$ is simply given by evaluating $\lw^t$ at $l = \ell$,
leading to:
\eq \label{effltree}
    \gamma^t[\ell] = \int d^4x \; \lw^t(\ell) .
\eeq

Similarly, evaluating the path integral expression of
eq.~\pref{wilsondef} to obtain $\lw$ in the tree approximation
entails evaluating the classical action at the saddle point $l =
\ell$ and $H = \hbr^t(\ell)$. Graphically, this gives the
tree-level Wilson action as the sum over all tree graphs which
have only heavy particles propagating in their internal lines, and
only light particles for external lines. Retracing the steps taken
in previous sections to compute $\gamma[\ell]$ at tree level gives
an explicit expression for $\lw^t = \Scl(\ell, \hbr^t(\ell))$:
\eqa \label{hereslw}
    \lw^t(\ell) &=& - \hf \; \partial_\mu \ell \, \partial^\mu \ell -
    \frac{m^2}{2} \,
    \ell^2 - \left( \frac{g_l}{4!}
    - \frac{\tilde{m}^2}{8M^2}\right) \, \ell^4  \\
    && \qquad - \left(  \frac{\tilde{m}^2}{2M^4} \right) \ell^2 \,
    \partial_\mu \ell \,
    \partial^\mu \ell -  \left( \frac{ g_{l h}
    \, \tilde{m}^2}{16M^4}  \right)
    \ell^6 + \cdots , \nn\\
    &=& - \hf \; \partial_\mu \ell \, \partial^\mu \ell -
    \frac{m^2}{2} \,
    \ell^2 - \left( \frac{g_l}{4!}
    - \frac{\tilde{m}^2}{8M^2} - \frac{m^2 \tilde{m}^2}{6M^4} \right) \, \ell^4  \\
    && \qquad -  \left( \frac{ g_{l h}
    \, \tilde{m}^2}{16M^4} - \frac{g_l \, \tilde{m}^2}{36 M^4} \right)
    \ell^6 + \cdots , \nn
\eeqa
where the freedom to redefine fields has been used, and ellipses
represent terms which are higher order in $1/M$ than are those
displayed. This result also could be obtained by asking what local
lagrangian involving only light fields reproduces the 2-body and
3-body scattering of the full theory to $O(1/M^4)$.

\subsubsection{One-Loop Calculation:}

At one loop many of the nontrivial features of the Wilson action
emerge for the first time. As usual we assume the renormalized
dimensional parameters which ensure the hierarchy of scales in the
two-scalar model satisfy $m \sim \tilde{m} \ll M$.

The most general possible Wilson action which is local and
consistent with the symmetry $\ell \to - \ell$ is
\eq \label{swguess}
    \sw[\ell] = -\int d^4x \; \left[a_0 + \frac{a_2}{2} \,\ell^2
    + \frac{a_4}{4!} \,\ell^4 + \frac{1}{2}(1 + b_2) \,\partial_\mu \ell
    \,\partial^\mu \ell + \cdots \right]\,,
\eeq
where ellipses denote higher-dimension interactions and the
constants $a_k$, $b_k$, \etc\ to be determined by matching to the
full theory, as is now described.

For simplicity we specialize to matching to $O(M^0)$, since this
allows us to specialize to background configurations for which
$\ell$ is spacetime independent --- \ie\ $\partial_\mu \ell = 0$.
The only effective couplings which are relevant in this case (more
about this in the next section) are $a_0$, $a_2$ and $a_4$, and it
is convenient to determine these by requiring the one-loop result
for $\gamma^\oneloop[\ell]$, computed using $\sw$, agrees with the
result computed with the full theory, to $O(M^0)$.

The previous section gives the one-loop calculation in the full
theory to this order as eq.~\pref{approxeffpotexpr}, evaluated at
$h = \hbr^t(\ell) = - (\tilde{m}/2M^2) \, \ell^2 + O(1/M^4)$. This
gives $\gamma^\oneloop_{\rm pot}[\ell] = - \int d^4x \;
V^\oneloop$, with $V^\oneloop$ as given in
eq.~\pref{oneloopfinpart}:
\eqa \label{oneloopfinpart1}
   V^\oneloop(\ell) &=&
    \nth{64 \pi^2} \left\{ M^4 \log \left( \frac{M^2}{\mu^2}
    \right)\right. \nn\\
    && \;\; + \,(g_{lh} M^2 - \tilde{g}_h \tilde{m}M + 2 \tilde{m}^2)
    \left[ \log \left( \frac{M^2}{\mu^2}
    \right) + \frac12 \right] \, \ell^2\nn\\
    && \quad + \frac{g_l^2}{4} \left[ \log \left( \frac{M^2}{\mu^2}
    \right) + \frac32 \right] \ell^4 \nn\\
    && \qquad + \left. \left( m^2 + \frac{g_l}{2} \, \ell^2
    \right)^2 \; \log\left( {m^2 + \frac{1}{2} \,g_l\,
    \ell^2 \over \mu^2} \right) \right\},
\eeqa
neglecting $O(1/M)$ terms. With ultraviolet divergences
regularized in $n = 4-2\varepsilon$ dimensions, and renormalized
using the $\msbar$ renormalization scheme, the counter-terms used
to obtain this expression are those of eq.~\pref{RG}:
\eqa \label{RG1}
    A_\ssr &=& A_0 - {1 \over 64 \pi^2} \; (M^4 + m^4) \; L , \nn\\
    B_\ssr &=& B_0 - {1 \over 32 \pi^2}\;
    (m^2 \tilde{m} + \tilde{g}_h M^3) \; L , \nn\\
    m^2_\ssr &=& m^2 - {1 \over 32 \pi^2}\;
    (g_l m^2 + g_{l h} M^2 + 2 \tilde{m}^2) \; L , \\
    M^2_\ssr &=& M^2 - {1 \over 32 \pi^2}\;   \Bigl(g_{l h} m^2 + (g_h
    + \tilde{g}_h^2) M^2 + \tilde{m}^2 \Bigr) \; L  , \nn\\
    \tilde{m}_\ssr &=& \tilde{m} - \, {1 \over 32 \pi^2} \; (g_l
    \tilde{m} + g_{l h} \tilde{g}_h M + 4 g_{l h} \tilde{m}) \; L , \nn\\
    (\tilde{g}_h)_\ssr &=& \tilde{g}_h - \, {3 \over 32 \pi^2} \;
    \left(g_{l h} {\tilde{m} \over M} +  g_h \tilde{g}_h \right) \; L  \nn\\
    (g_l)_\ssr &=& g_l - \, {3 \over 32 \pi^2} \; (g_l^2 + g_{l
    h}^2) \; L \nn\\
    (g_h)_\ssr &=& g_h  - \, {3 \over 32 \pi^2} \;  (g_h^2 + g_{l
    h}^2) \; L \nn\\
    (g_{l h})_\ssr &=& g_{l h} - \, {1 \over 32 \pi^2} \;
    (g_l + g_h +  4 g_{l h}) g_{l h} \; L , \nn
\eeqa
with $L = 1/\varepsilon + k$.

Repeating the same calculation using the Wilson action,
eq.~\pref{swguess}, instead leads to:
\eq \label{wilsononeloop}
    V^\oneloop_{\rm eff} = \nth{64 \pi^2} \,
    \left( \hat{a}_2 + \frac{\hat{a}_4}{2} \, \ell^2 \right)^2
    \log\left( { \hat{a}_2 + \frac{\hat{a}_4}{2} \, \ell^2
    \over \mu^2} \right),
\eeq
where $\hat{a}_2 = a_2/(1 + b_2) \approx a_2 (1 - b_2 + \cdots)$
and $\hat{a}_4 = a_4/(1 + b_2)^2 \approx a_4 (1 - 2 b_2 +
\cdots)$. Here the constants $a_k$ have been renormalized in the
effective theory, being related to the bare couplings by
expressions similar to eq.~\pref{RG1}:
\eqa \label{RG2}
    (a_0)_\ssr &=& a_0 - {a_2^2 \over 64 \pi^2} \; L , \nn\\
    (a_2)_\ssr &=& a_2 - {a_4 a_2 \over 32 \pi^2}\;L , \\
    (a_4)_\ssr &=& a_4 - \, {3 \, a_4^2 \over 32 \pi^2} \; L .\nn
\eeqa

The constants $a_k$ are now fixed by performing the UV-finite
renormalization required to make the two calculations of $V^t +
V^\oneloop$ agree. The $b_k$ are similarly determined by matching
the two-derivative terms in $\gamma^\oneloop$, and so on. (For the
scalar model under consideration this gives $b_2 = 0$ to the order
of interest, since a single $H$ loop first contributes to the
$\partial_\mu \ell \, \partial^\mu \ell$ term at $O(1/M^2)$.)
Requiring the tree contribution, $V^t_{\rm eff} = a_0 + \frac12
a_2 \, \ell^2 + \frac{1}{4!} \, a_4 \, \ell^4$, to capture the
terms in $V^\oneloop$ which are missing in $V^\oneloop_{\rm eff}$
gives the required effective couplings in the Wilson action:
\eqa \label{matching1}
    (a_0)'_\ssr &=& A + \nth{64 \pi^2} \, M^4 \log \left( \frac{M^2}{\mu^2}
    \right) \nn\\
    (a_2)'_\ssr &=& m^2 + \nth{64 \pi^2}  \,(g_{lh} M^2 - \tilde{g}_h \tilde{m}M
    + 2 \tilde{m}^2) \left[2 \log \left( \frac{M^2}{\mu^2}
    \right) + 1 \right]  \nn\\
    (a_4)'_\ssr &=& g_l + \frac{3\, g_l^2}{32 \pi^2}  \,
    \left[ \log \left( \frac{M^2}{\mu^2}
    \right) + \frac32 \right]  .
\eeqa

Such finite renormalizations, arising as a heavy particle is
integrated out, are called {\em threshold corrections}, and it is
through these that the explicit powers of $M$ get into the
low-energy theory in dimensional regularization. (Notice also that
the coefficients of $M^2$ in these expressions need not agree with
those of $\Lambda^2$ in a cutoff low-energy theory, revealing the
fallacy of using quadratic divergences in an effective theory to
track heavy-mass dependence \cite{CutoffCaveat}.)

\section{Power counting}

As is clear from the two-scalar model, effective lagrangians
typically involve potentially an infinite number of interactions
corresponding to the ultimately infinite numbers of terms which
can arise once the $M$-dependence of physical observables is
expanded in powers of $1/M$. If it were necessary to deal with
even a large number of these terms there would be no real utility
in using them in practical calculations.

The most important part of an effective-lagrangian analysis is
therefore the identification of which terms in $\leff$ are
required in order to compute observables to any given order in
$1/M$, and this is accomplished using the power-counting rules of
this section.

\subsection{A class of effective interactions}

To keep the discussion interestingly general, in this section we
focus on a broad class of effective lagrangians that can be
written in the following way:
\eq \label{leffpc}
    \leff = f^4 \sum_k {c_k \over M^{d_k}} \;
    \Sco_k \left( {\phi \over v} \right).
\eeq
In this expression, $\phi$ is meant to generically represent the
fields of the problem, which for simplicity of presentation are
taken here to be bosons and so to have the canonical dimension of
mass, using fundamental units for which $\hbar = c = 1$. (The
generalization to more general situations, including fermions, is
straightforward.) The quantities $f$, $v$ and $M$ are all
constants also having the dimensions of mass. The index `$k$' runs
over all of the labels of the various effective interactions which
appear in $\leff$, and which are denoted by $\Sco_k$, and which
are assumed to have dimension (mass)${}^{d_k}$. Since the ratio
$\phi/v$ is dimensionless, all of this dimension is carried by
derivatives, $\partial \phi$. As a result, $d_k$ simply counts the
number of derivatives which appear in the effective interaction,
$\Sco_k$.

\subsection{Power-counting rules}

Imagine now computing Feynman graphs using these effective
interactions, with the goal of tracking how the result depends on
the scales $f$, $v$ and $M$, as well as the mass scale, $m$, of
the low-energy particles. Consider in particular a graph,
$\Sca_\sse(q)$, involving $E$ external lines whose four-momenta
are collectively denoted by $q$. Suppose also that this graph has
$I$ internal lines and $V_{ik}$ vertices. The labels $i$ and $k$
indicate two of the properties of the vertices: with $i$ counting
the number of lines which converge at the vertex, and $k$ counting
the power of momentum which appears in the vertex. Equivalently,
$i$ counts the number of powers of the fields, $\phi$, which
appear in the corresponding interaction term in the lagrangian,
and $k$ counts the number of derivatives of these fields which
appear there.

\subsubsection{Some useful identities:}

The positive integers, $I$, $E$ and $V_{ik}$, which characterize
the Feynman graph in question are not all independent since they
are related by the rules for constructing graphs from lines and
vertices. One such a relation is obtained by equating the two
equivalent ways of counting the number of ends of internal and
external lines in a graph. On one hand, since all lines end at a
vertex, the number of ends is given by summing over all of the
ends which appear in all of the vertices: $\sum_{ik} i \, V_{ik}$.
On the other hand, there are two ends for each internal line, and
one end for each external line in the graph: $2 I + E$. Equating
these gives the identity which expresses the `conservation of
ends':
\eq \label{consofends}
    2 I + E = \sum_{ik} i \,  V_{ik}, \qquad
    \hbox{(Conservation of Ends)}.
\eeq

A second useful identity {\em defines} of the number of loops,
$L$, for each (connected) graph:
\eq \label{loopdef}
    L = 1 + I - \sum_{ik} V_{ik}, \qquad
    \hbox{(Definition of $L$)}.
\eeq
This definition doesn't come out of thin air, since for graphs
which can be drawn on a plane it agrees with the intuitive notion
of the number of loops in a graph.

\subsubsection{Estimating integrals:}

Reading the Feynman rules from the lagrangian of eq.~\pref{leffpc}
shows that the vertices in the Feynman graph of interest
contribute a factor
\eq \label{vertexcont}
    \hbox{(Vertex)} =  \prod_{ik} \left[ i (2
    \pi)^4 \delta^4(p) \; \left( {p \over M} \right)^k \; \left( {f^4
    \over v^i} \right) \right]^{V_{ik}},
\eeq
where $p$ generically denotes the various momenta running through
the vertex.

Similarly, each internal line contributes the additional factors:
\eq \label{internallinecont}
    \hbox{(Internal Line)} = \left[ -i
    \int {d^4 p \over (2 \pi)^4} \; \left( {M^2 v^2 \over
    f^4} \right) \; {1 \over p^2 + m^2} \right]^I,
\eeq
where, again, $p$ denotes the generic momentum flowing through the
line. $m$ denotes the mass of the light particles which appear in
the effective theory, and it is assumed that the kinetic terms
which define their propagation are those terms in $\leff$
involving two derivatives and two powers of the fields, $\phi$.

As usual for a connected graph, all but one of the
momentum-conserving delta functions in eq.~\pref{vertexcont} can
be used to perform one of the momentum integrals in
eq.~\pref{internallinecont}. The one remaining delta function
which is left after doing so depends only on the external momenta,
$\delta^4(q)$, and expresses the overall conservation of
four-momentum for the process. Future formulae are less cluttered
if this factor is extracted once and for all, by defining the
reduced amplitude, $A$, by
\eq \label{redampdef} \Sca_\sse(q) = i (2 \pi)^4 \delta^4(q) \;
A_\sse(q). \eeq

The number of four-momentum integrations which are left after
having used all of the momentum-conserving delta functions is then
$I - \sum_{ik} V_{ik} + 1 = L$. This last equality uses the
definition, eq.~\pref{loopdef}, of the number of loops, $L$.

In order to track how the result depends on the scales in $\leff$
it is convenient to estimate the results of performing the various
multi-dimensional momentum integrals using dimensional analysis.
Since these integrals are typically ultraviolet divergent, they
must first be regulated, and this is where the use of dimensional
regularization pays off. The key observation is that if a
dimensionally regulated integral has dimensions, then its size is
set by the light masses or external momenta which appear in the
integrand. That is, dimensional analysis applied to a
dimensionally-regulated integral implies
\eq \label{newdimgrounds}
    \int \cdots \int \left( {d^n p\over (2
    \pi)^n} \right)^A \; {p^B \over (p^2 + m^2)^C }  \sim \left( {1
    \over 4 \pi} \right)^{2A} m^{nA + B - 2C} ,
\eeq
with a dimensionless prefactor which depends on the dimension,
$n$, of spacetime, and which may be singular in the limit that $n
\to 4$. Here $m$ represents the dominant scale which appears in
the integrand of the momentum integrations. If the light particles
appearing as external states in $A_\sse(q)$ should be massless, or
highly relativistic, then the typical external momenta, $q$, are
much larger than $m$, and $m$ in the above expression should be
replaced by $q$.\footnote{Any logarithmic infrared mass
singularities which may arise in this limit are ignored here,
since our interest is in following {\em powers} of ratios of the
light and heavy mass scales.} $q$ is used as the light scale
controlling the size of the momentum integrations in the formulae
quoted below.

With this estimate for the size of the momentum integrations, we
find the following quantity appears in the amplitude $A_\sse(q)$
\eq \label{intcontribution}
    \int \cdots \int \left( {d^4 p\over (2
    \pi)^4} \right)^L \; {p^{\sum_{ik} k V_{ik}} \over (p^2 + q^2)^I }
    \sim \left( {1 \over 4 \pi} \right)^{2L} q^{4L - 2I + \sum_{ik} k
    V_{ik}} ,
\eeq
which, with liberal use of the identities \pref{consofends} and
\pref{loopdef}, gives as estimate for $A_\sse(q)$:
\eq \label{aedwdimreg}
    A_\sse(q) \sim f^4 \; \left( {1 \over
    v} \right)^E \; \left( {Mq \over 4 \pi f^2} \right)^{2L} \;
    \left( {q \over M} \right)^{2 + \sum_{ik} (k - 2) V_{ik}} .
\eeq

This last formula is the main result, which is used in the various
applications considered later. Its utility lies in the fact that
it links the contributions of the various effective interactions
in the effective lagrangian, \pref{leffpc}, with the dependence of
observables on small mass ratios such as $q/M$. Notice in
particular that more and more complicated graphs -- for which $L$
and $V_{ik}$ become larger and larger -- are only suppressed in
their contributions to observables if $q$ is much smaller than the
scales $M$ and $f$.

Notice also that the basic estimate, eq.~\pref{newdimgrounds},
would have been much more difficult to do if the effective
couplings were defined using a cutoff, $\lambda$, because in this
case it is the cutoff which would dominate the integral, since it
is then the largest external scale. But we knowing this cutoff
dependence is less useful because the general arguments of the
previous sections show that $\lambda$ is guaranteed to drop out of
any physical quantity.

\subsection{The Effective-Lagrangian Logic}

Power-counting estimates of this sort suggest the following
general logic concerning the use of effective lagrangians:

\begin{description}

\item[Step I] Choose the accuracy (\eg\ one part per mille) with
which observables, such as $A_\sse(q)$, are to be computed.

\item[Step II] Determine the order in the small mass ratios $q/M$
or $m/M$ that must be required in order to achieve the desired
accuracy.

\item[Step III] Use the power counting result,
eq.~\pref{aedwdimreg}, to find which terms in the effective
lagrangian are needed in order to compute to the desired order in
$q/M$ and $m/M$. Eq.~\pref{aedwdimreg} also determines which order
in the loop expansion is required for each effective interaction
of interest.

\item[Step IVa] Compute the couplings of the required effective
interactions using the full underlying theory. If this step should
prove to be impossible, due either to ignorance of the underlying
theory or to the intractability of the required calculation, then
it may be replaced by the following alternative:

\item[Step IVb] If the coefficients of the required terms in the
effective lagrangian cannot be computed then they may instead be
regarded as unknown parameters which are to be taken from
experiment. Once a sufficient number of observables are used to
determine these parameters, all other observables may be
unambiguously predicted using the effective theory.

\end{description}

A number of points cry out for comment at this point.

\smallskip\noindent$\bullet$ {Utility of Step IVb:}
The possibility of treating the effective lagrangian
phenomenologically, as in Step IVb above, immeasurably broadens
the utility of effective lagrangian techniques, since they need
not be restricted to situations for which the underlying theory is
both known and calculationally simple. Implicit in such a program
is the underlying assumption that there is no loss of generality
in working with a local field theory. This assumption has been
borne out in all known examples of physical systems. It is based
on the conviction that the restrictions which are implicit in
working with local field theories are simply those that follow
from general physical principles, such as unitarity and cluster
decomposition.

\smallskip\noindent$\bullet$ {\it When to expect renormalizability:}
Since eq.~\pref{aedwdimreg} states that only a finite number of
terms in $\leff$ contribute to any fixed order in $q/M$, and these
terms need appear in only a finite number of loops, it follows
that only a finite amount of labour is required to obtain a fixed
accuracy in observables. Renormalizable theories represent the
special case for which it suffices to work only to zeroth order in
the ratio $q/M$. This can be thought of as being the reason why
renormalizable theories play such an important role throughout
physics.

\smallskip\noindent$\bullet$ {\it How to predict using nonrenormalizable
theories:} An interesting corollary of the above observations is
the fact that only a finite number of renormalizations are
required in the low-energy theory in order to make finite the
predictions for observables to any fixed order in $q/M$. Thus,
although an effective lagrangian is not renormalizable in the
traditional sense, it nevertheless {\em is} predictive in the same
way that a renormalizable theory is.

\section{Applications}

We next turn briefly to a few illustrative applications of these
techniques.

\subsection{Quantum Electrodynamics}

The very lightest electromagnetically interacting elementary
particles are the photon and the electron, and from the general
arguments given above we expect that the effective field theory
which describes their dominant low-energy interactions should be
renormalizable, corresponding to the neglect of all inverse masses
heavier than the electron. The most general such renormalizable
interactions which are possible given the electron charge
assignment is
\eq \label{QEDlagrangian}
    \Scl_{\rm QED}(A_\mu,\psi) = - \, \nth{4} \, F_{\mu\nu} F^{\mu\nu}
    - \psibr (\Dslsh + m_e) \psi ,
\eeq
where $F_{\mu\nu} = \partial_\mu A_\nu - \partial_\nu A_\mu$ gives
the electromagnetic field strength in terms of the electromagnetic
potential, $A_\mu$, and the electron field, $\psi$, is a
(four-component) Dirac spinor. The covariant derivative for the
electron field is defined by $D_\mu \psi =
\partial_\mu \psi + ie A_\mu \psi$, where $e$ is the
electromagnetic coupling constant.

Notice that this is precisely the lagrangian of Quantum
electrodynamics (QED). In this we have the roots of an explanation
of {\em why} QED is such a successful description of
electron-photon interactions.

\subsubsection{Integrating Out the Electron:}

Many practical applications of electromagnetism involve the
interaction of photons with macroscopic electric charge and
current distributions at energies, $E$, and momenta, $p$, much
smaller than the electron mass, $m_e$. As a result they fall
within the purview of low-energy techniques, and so lend
themselves to being described by an effective theory which is
defined below the electron mass, $m_e$, as is now described.

For present purposes it suffices to describe the macroscopic
charge distributions using an external electromagnetic current,
$J^\mu_{\rm em}$, which can be considered as an approximate,
mean-field, description of a collection of electrons in a real
material. This approximation is extremely good for macroscopically
large systems, if these systems are only probed by electromagnetic
fields whose energies are very small compared to their typical
electronic energies. We take the interaction term coupling the
electromagnetic field to this current to be the lowest dimension
interaction that is possible:
\eq \label{QEDJterm}
    \Scl_J = - e \, A_\mu \; J_{\rm em}^\mu .
\eeq
This coupling is only consistent with electromagnetic gauge
invariance if the external current is identically conserved:
$\partial_\mu J_{\rm em}^\mu = 0$, independent of any equations of
motion, and falls off sufficiently quickly to ensure that there is
no current flow at spatial infinity. Two practical examples of
such conserved configurations would be those of ($i$) a static
charge distribution: $J^0_{\rm em} = \rho(\bfr)$, $\Bfj_{\rm em} =
0$, or ($ii$) a static electrical current: $J^0_{\rm em} = 0$ and
$\Bfj_{\rm em} = \bfj(\bfr)$, where $\rho(\bfr)$ and $\bfj(\bfr)$
are localized, time-independent distributions, satisfying $\nabla
\cdot \bfj = 0$.

Our interest is in the properties of electromagnetic fields
outside of such distributions, and of tracking in particular the
low-energy effects of virtual electrons. The most general
effective theory involving photons only which can govern the
low-energy limit is:
\eq \label{totQEDefflagr}
    \Scl_4 = \leff (A)- e \, A_\mu \; J_{\rm em}^\mu .
\eeq
where $\leff$ may be expanded in terms of interactions having
successively higher dimensions. Writing $\leff = \Scl_4 + \Scl_6 +
\Scl_8 + \cdots$ we have
\eqa \label{genQEDefflagr}
    \Scl_4 &=& - \, {Z \over 4} \; F_{\mu\nu} F^{\mu\nu}, \\
    \Scl_6 &=& {a \over m_e^2} \, F_{\mu\nu} \Box F^{\mu\nu} + {a'
    \over
    m_e^2} \, \partial_\mu F^{\mu\nu} \partial^\lambda F_{\lambda\nu} , \nn\\
    \Scl_8 &=& {b \over m_e^4} \, (F_{\mu\nu} F^{\mu\nu})^2 + {c \over
    m_e^4} \, (F_{\mu\nu} \tilde{F}^{\mu\nu})^2  + \hbox{($\partial^4
    F^2$ terms)},
\eeqa
and so on. In this expression for $\Scl_8$, $\tilde{F}$ represents
the `dual' field-strength tensor, defined by $\tilde{F}_{\mu\nu} =
\hf \, \epsilon_{\mu\nu\lambda\rho} F^{\lambda\rho}$, and possible
terms involving more derivatives exist, but have not been written.
A power of $1/m_e$ has been made explicit in the coefficient of
each term, since we work perturbatively in the electromagnetic
coupling, and $m_e$ is the only mass scale which appears in the
underlying QED lagrangian. With this power extracted, the
remaining constants, $Z$, $a$, $a'$, $b$, $c$, \etc, are
dimensionless.

Notice that no $\partial^2 F^3$ terms appear since any terms
involving an odd number of $F$'s are forbidden by
charge-conjugation invariance, for which $F_{\mu\nu} \to -
F_{\mu\nu}$. This is Furry's theorem in its modern, low-energy,
guise.

Notice also that we may use the freedom to redefine fields to
rewrite the $1/m_e^2$ terms in terms of direct current-current
`contact' interactions. The simplest way to do so is to rewrite
these interactions using the lowest-dimension equations of motion,
$\partial_\mu F^{\mu\nu} = e J_{\rm em}^\nu$ -- and so also $\Box
F^{\mu\nu} = e (\partial^\mu J_{\rm em}^\nu - \partial^\nu J_{\rm
em}^\mu)$ -- following the general arguments of earlier sections.
Together with an integration by parts, this allows $\Scl_6$ to be
written
\eq \label{genQEDefflagr1}
    \Scl_6 = {e^2 \over m_e^2} \, (a' - 2a) \,
    J_{\rm em}^\nu J_{{\rm em}\,\nu} .
\eeq
This shows in particular that these interactions are irrelevant
for photon propagation and scattering.\footnote{Things are more
interesting if conducting boundaries are present however, see
\cite{Boundaries}.}

In principle, we may now proceed to determine which of these
effective interactions are required when working to a fixed order
in $1/m_e^2$ in any given physical observable. Having done so, we
may then compute the relevant dimensionless coefficients $Z$, $a'
- 2a$, $b$, $c$ \etc. and thereby retrieve the low-energy limit of
the full QED prediction for the observable in question.

\subsubsection{The Scattering of Light by Light:}

Inspection of the effective lagrangian, eq.~\pref{genQEDefflagr},
shows that the simplest interaction to receive contributions to
$\Sco(m_e^{-4})$ is the scattering of light by light, for which
the leading contribution to the cross section at centre-of-mass
energies $E_{\rm cm} \ll m_e$ can be inferred relatively easily.

The first step is to determine precisely which Feynman graphs
built from the interactions in the effective lagrangian of
eq.~\pref{genQEDefflagr} contribute, order-by-order in $1/m_e$.
For this we may directly use the general power-counting results of
the previous chapter, since the effective lagrangian of
eq.~\pref{genQEDefflagr} is a special case of the form considered
in eq.~\pref{leffpc}, with the appropriate dimensionful constants
being $f = M = v = m_e$. Directly using eq.~\pref{aedwdimreg} for
the $E$-point scattering amplitude, $A_\sse(q)$, leads in the
present case to:
\eq \label{QEDpc}
    A_\sse(q) \sim q^2 m_e^2 \; \left( {1 \over
    m_e} \right)^E \; \left( {q \over 4 \pi \, m_e} \right)^{2L} \; \left( {q
    \over m_e} \right)^{\sum_{ik} (k - 2) V_{ik}} .
\eeq

We may also use some specific information for the QED lagrangian
which follows from the gauge invariance of the problem. In
particular, since the gauge potential, $A_\mu$, only appears in
$\leff$ through its field strength, $F_{\mu\nu}$, all of the
interactions of the effective theory must contain at least as many
derivatives as they have powers of $A_\mu$. In equations: $V_{ik}
= 0$ unless $k \geq i$. In particular, since $k=2$ therefore
implies $i \leq 2$, we see that the only term in $\leff$ having
exactly two derivatives is the kinetic term, $F_{\mu\nu}
F^{\mu\nu}$. Since this is purely quadratic in $A_\mu$, it is not
an interaction (rather, it is the unperturbed lagrangian), and so
we may take $V_{ik} = 0$, for $k \leq 2$. A consequence of these
considerations is the inequality $\sum_{ik} (k - 2) V_{ik} \geq
2$, and this sum equals 2 only if $V_{ik} = 0$ for all $k > 4$,
and if $V_{i4} = 1$.

\begin{figure}
\setlength{\unitlength}{1mm}
\centerline{%
\begin{picture}(50,30)
    \thicklines
    \multiput(29,21)(16,0){2}{\oval(8,8)[b]}
    \multiput(37,21)(16,0){1}{\oval(8,8)[t]}
    \multiput(29,21)(0,16){2}{\oval(8,8)[l]}
    \multiput(29,29)(0,16){1}{\oval(8,8)[r]}
    \put(25,21){\circle*{12}}
    \multiput(5,21)(16,0){2}{\oval(8,8)[t]}
    \multiput(13,21)(16,0){1}{\oval(8,8)[b]}
    \multiput(21,5)(0,16){2}{\oval(8,8)[r]}
    \multiput(21,13)(0,16){1}{\oval(8,8)[l]}
\end{picture}}
\caption{The Feynman graph which contributes the leading
contribution to photon-photon scattering in the effective theory
for low-energy QED. The vertex represents either of the two
dimension-eight interactions discussed in the text.}
\label{Fgraphphoton}
\end{figure}
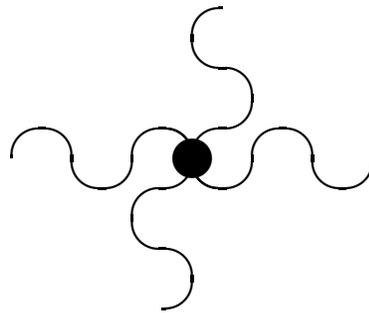

For two-body photon-photon scattering we may take $E=4$, and from
eq.~\pref{QEDpc} it is clear that the minimum power of $q/m_e$
which can appear in $A_4(q)$ corresponds to taking $(i)$ $L=0$,
and $(ii)$ $V_{ik} = 0$ for $k \ne 4$, and $V_{i4} = 1$ for
precisely one vertex for which $k=4$. This tells us that the only
graph which is relevant for photon-photon scattering at leading
order in $E_{\rm cm}/m_e$ is the one shown in
Fig.~\pref{Fgraphphoton}, which uses precisely one of the two
vertices in $\Scl^{(8)}$, for which $i=k=4$, and gives a result of
order $q^6/m_e^8$ (where $q \sim E_{\rm cm}$).

The effective lagrangian really starts saving work when
sub-leading contributions are computed. For photon-photon
scattering eq.~\pref{QEDpc} implies the terms suppressed by two
additional powers of $E_{\rm cm}/m_e$ requires one insertion of
the $k=4, i=2$ interaction, but since this is in $\Scl_6$ it
cannot contribute, making the next-to-leading contribution down by
at least $(E_{\rm cm}/m_e)^4$ relative to the leading terms.

The leading term comes from using the quartic interactions of
$\Scl_8$ in eq.~\pref{genQEDefflagr} in the Born approximation.
This gives the following differential cross section for
unpolarized photon scattering, in the CM frame:
\eq \label{ggscattering}
    {d \sigma_{\gamma\gamma} \over d \Omega}
    = {278 \over 65 \pi^2} \; \left[(b+c)^2 + (b-c)^2 \right] \;
    \left( {E_{\rm cm}^6 \over m_e^8} \right) \left( 3 + \cos^2 \theta
    \right)^2 \; \left[ 1 + \Sco\left( {E_{\rm cm}^4 \over m_e^4} \right)
    \right] .
\eeq
Here $E_{\rm cm}$ is the energy of either photon in the CM frame,
$d\Omega$ is the differential element of solid angle for one of
the outgoing photons, and $\theta$ is the angular position of this
solid-angle element relative to the direction of (either of) the
incoming photons. Notice that, to this point, we have used only
the expansion in powers of $E_{\rm cm}/m_e$, and have {\em not}
also expanded the cross section in powers of $\alpha = e^2/4\pi$.

\begin{figure}
\setlength{\unitlength}{1mm}
\centerline{%
\begin{picture}(50,30)
    \thicklines
    \multiput(49,11)(16,0){2}{\oval(8,8)[b]}
    \multiput(57,11)(16,0){1}{\oval(8,8)[t]}
    \multiput(49,51)(16,0){2}{\oval(8,8)[t]}
    \multiput(57,51)(16,0){1}{\oval(8,8)[b]}
    \put(45,51){\circle*{4}}
    \put(45,51){\line(0,-1){40}}
    \put(45,51){\vector(0,-1){24}}
    \put(45,10){\circle*{4}}
    \multiput(-8,11)(16,0){1}{\oval(8,8)[t]}
    \multiput(-16,11)(16,0){2}{\oval(8,8)[b]}
    \multiput(-8,51)(16,0){1}{\oval(8,8)[b]}
    \multiput(-16,51)(16,0){2}{\oval(8,8)[t]}
    \put(5,51){\circle*{4}}
    \put(5,51){\line(1,0){40}}
    \put(6,51){\vector(1,0){20}}
    \put(5,10){\line(0,1){40}}
    \put(5,10){\vector(0,1){24}}
    \put(5,10){\circle*{4}}
    \put(5,10){\line(1,0){40}}
    \put(45,10){\vector(-1,0){20}}
\end{picture}}
\caption{The leading Feynman graphs in QED which generate the
effective four-photon operators in the low energy theory. Straight
(wavy) lines represent electrons (photons).} \label{FgraphBox}
\end{figure}
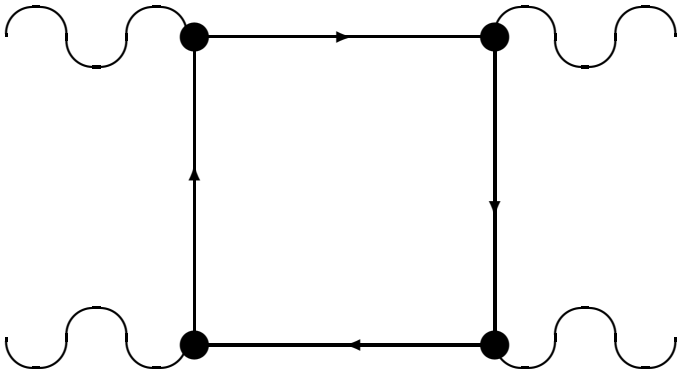

The final result for QED is reproduced once the constants $b$ and
$c$ are computed from the underlying theory, and it is at this
point that we first appeal to perturbation theory in $\alpha$. To
lowest order in $\alpha$, the relevant graph is shown in
Fig.~\pref{FgraphBox}. This graph is ultraviolet finite so no
counterterms are required, and the result is simply:
\eq \label{QEDcoefs}
    b = {4 \over 7} \; c = {\alpha^2 \over 90} ,
\eeq
leading to the standard result \cite{PhotonScattering}
\eq \label{ggscattering1}
    {d \sigma_{\gamma\gamma} \over d \Omega}
    = {139 \over 4\pi^2} \; \left( \frac{\alpha^2}{90} \right)^2  \;
    \left( {E_{\rm cm}^6 \over m_e^8} \right) \left( 3 + \cos^2 \theta
    \right)^2 \; \left[ 1 + \Sco\left( {E_{\rm cm}^4 \over m_e^4} \right)
    \right].
\eeq

\subsubsection{Renormalization and Large Logs:}

We next return to the terms in $\leff$ which are unsuppressed at
low energies by powers of $1/m_e$, for both practical and
pedagogical reasons. Pedagogy is served by using this to introduce
the {\em Decoupling Subtraction} renormalization scheme, which is
the natural generalization of minimal subtraction to the effective
lagrangian framework. The practical purpose of this example is to
demonstrate how classical electromagnetism gives an {\em exact}
description of photon response at low energies, whose corrections
require powers of $1/m_e$ (and {\em not}, for example, simply more
powers of $\alpha$).

The only term in the effective lagrangian of
eq.~\pref{genQEDefflagr} which is not suppressed by powers of
$1/m_e$ is the term
\eq \label{totQEDefflagr}
    \Scl_4 = - \, {Z \over 4} \; F_{\mu\nu}
    F^{\mu\nu}  - e \, A_\mu \; J_{\rm em}^\mu .
\eeq
All of the influence of the underlying physics appears here only
through the dimensionless parameter, $Z$, whose leading
contribution from virtual electrons may be explicitly computed in
QED to be
\eq \label{Zeonly}
    Z = 1 - {\alpha \over 3 \pi} \left[ \nth{\veps}
    + k + \log \left( {m_e^2 \over \mu^2} \right) \right],
\eeq
where we regulate ultraviolet divergences using dimensional
regularization. $k$ is the constant encountered in section
\ref{dimregWilson}, which appears universally with the divergence,
$1/\veps$, in dimensional regularization, and $\mu$ is the usual
(arbitrary) mass scale introduced in dimensional regularization to
keep the coupling constant, $e$, dimensionless.

The physical interpretation of $Z$ is found by performing the
rescaling $A_\mu = Z^{-\hf} \; A_\mu^\ssr$, to put the photon
kinetic term into canonical form. In this case we recover the
effective theory
\eq \label{totQEDefflagr1}
    \Scl_4 = - \, {1 \over 4} \; F_{\mu\nu}
    F^{\mu\nu}  - e_{\rm phys} \, A_\mu \; J_{\rm em}^\mu .
\eeq
where $e \equiv Z^{\hf} \; e_{\rm phys}$. We see that to leading
order virtual electrons affect low-energy photon properties only
through the {\em value} taken by the physical electric charge,
$e_{\rm phys}$, and does not otherwise affect electromagnetic
properties.

This conclusion has important practical implications concerning
the accuracy of the calculations of electromagnetic properties at
low energies in QED. It states that the justification of simply
using Maxwell's equations to describe photon properties at low
energies is the neglect of terms of order $1/m_e$, and not the
neglect of powers of $\alpha$. Thus, for example, even though the
classical formulae for the scattering of electromagnetic waves by
a given charge distribution is a result which is obtained only in
tree approximation in QED, corrections to these formulae do not
arise at any order in $\alpha$ unsuppressed by powers of $E_{\rm
cm}/m_e$. To zeroth order in $1/m_e$ the sole affect of all
higher-loop corrections to electromagnetic scattering is to
renormalize the value of $\alpha$ in terms of which all
observables are computed. The significance of this renormalization
enters once it is possible to measure the coupling at more than
one scale, since then the logarithmic running of couplings with
scale causes real physical effects, in particular encoding the
potential dependence of observables on some of the logarithms of
large mass ratios.

To see how to extract large logarithms most efficiently, in this
section we first contrasting two useful renormalization schemes.
The first of these is the one defined above, in which all of $Z$
is completely absorbed into the fields and couplings:
\eqa \label{physscheme}
    &&A_\mu = Z^{-\hf}_{\rm phys} \;
    A_\mu^{\rm phys}, \qquad
    \hbox{and} \qquad e = Z^{\hf}_{\rm phys} \; e_{\rm phys} , \nn\\
    \hbox{with} \qquad &&Z_{\rm phys} = Z = 1 - {\alpha \over 3 \pi}
    \left[ \nth{\veps} + k + \log \left( {m_e^2 \over \mu^2} \right)
    \right].
\eeqa
The subscript `phys' emphasizes that the charge, $e_{\rm phys}$,
is a physical observable whose value can be experimentally
determined, in principle. For instance, it could be measured by
taking a known static charge distribution, containing a
predetermined number of electrons, and then using Maxwell's
equations to predict the resulting flux of electric field at large
distances from these charges. Comparing this calculated flux with
the measured flux gives a measurement of $e_{\rm phys}$.

The alternative scheme of choice for most practical calculations
is the $\msbar$ scheme, for which the renormalization is defined
to subtract only the term $1/\veps + k$ in $Z$. That is:
\eqa \label{msbarscheme}
    &&A_\mu = Z^{-\hf}_\msbar \; A_\mu^\msbar
    , \qquad
    \hbox{and} \qquad e = Z^{\hf}_\msbar \; e_\msbar  , \nn\\
    &&\hbox{with} \qquad Z_\msbar = 1 - {\alpha \over 3 \pi} \left[
    \nth{\veps} + k \right].
\eeqa
In terms of this scheme the effective lagrangian becomes, to this
order in $\alpha$:
\eq \label{leffmsbar}
    \leff = - \, \nth{4} \left[ 1 - {\alpha
    \over 3 \pi} \log \left( {m_e^2 \over \mu^2} \right) \right] \;
    F_{\mu\nu}^\msbar F^{\mu\nu}_\msbar  - e_\msbar \, A_\mu^\msbar \;
    J_{\rm em}^\mu .
\eeq
The $\msbar$ coupling defined in this way is not itself a physical
quantity, but is simply a parameter in terms of which the
effective lagrangian is expressed. The physical coupling
$\alpha_{\rm phys} = e^2_{\rm phys}/4\pi$ and the $\msbar$
coupling $\alpha_\msbar = e^2_\msbar/4 \pi$ are related in the
following way:
\eq \label{schemerelation}
    \alpha_\msbar = \left( {Z_{\rm phys}
    \over Z_\msbar} \right) \;
    \alpha_{\rm phys}
    = \left[ 1 - {\alpha \over 3 \pi} \log \left( {m_e^2 \over
    \mu^2} \right) \right]\; \alpha_{\rm phys} .
\eeq

A key observation is that, since $\alpha_{\rm phys}$ is a physical
quantity, it cannot depend on the arbitrary scale $\mu$. As a
result, this last equation implies a $\mu$-dependence for
$\alpha_\msbar$. It is only for $\mu = m_e$ that the two couplings
agree:
\eq \label{rgbc} \alpha_\msbar(\mu = m_e) = \alpha_{\rm phys}.
\eeq

As usual, there is profit to be gained by re-expressing the $\mu$
dependence of $\alpha_\msbar$ of eq.~\pref{schemerelation} as a
differential relation:
\eq \label{rgeqtn}
    \mu^2 {d \alpha_\msbar \over d \mu^2 } =  + \;
    {\alpha^2_\msbar \over 3 \pi}.
\eeq
This is useful because the differential expression applies so long
as $\alpha_\msbar \ll 1$, while eq.~\pref{schemerelation} also
requires $\alpha_\msbar \log(m_e^2/\mu^2) \ll 1$. Integrating
eq.~\pref{rgeqtn} allows a broader inference of the
$\mu$-dependence of $\alpha_\msbar$:
\eq \label{rgimproved}
    {1 \over \alpha_\msbar(\mu)} = {1 \over
    \alpha_\msbar(\mu_0)} - {1 \over 3 \pi} \; \log\left( {\mu^2 \over
    \mu_0^2 } \right),
\eeq
which is accurate to all orders in $(\alpha/3 \pi) \log( m_e^2 /
\mu^2 )$, so long as $(\alpha/3 \pi) \ll 1$.

Eq.~\pref{rgimproved} is useful because it provides a simple way
to keep track of how some large logarithms appear in physical
observables. For instance, consider the cross section for the
scattering of electrons (plus an indeterminate number of soft
photons, having energies up to $E_{\rm max} = f E$) with $1 > f
\gg m_e/E$). Such a quantity has a smooth limit as $m_e/E \to 0$
when it is expressed in terms of $\alpha_\msbar(\mu)$
\cite{RGGoodThing,RenText}, so on dimensional grounds we may write
\eq \label{RGobs}
    \sigma(E,m_e,\alpha_{\rm phys}) = \frac{1}{E^2} \;\left[
    \Scf\left( \frac{E}{\mu} , \alpha_{\msbar}(\mu), f, \theta_k \right)
    + O\left( \frac{m_e}{E} \right) \right],
\eeq
where the $\theta_k$ denote any number of dimensionless quantities
(like angles) on which the observable depends, and the explicit
$\mu$-dependence of $\Scf$ must cancel the $\mu$-dependence which
appears implicitly through $\alpha_\msbar(\mu)$. However $\Scf$
{\em is} singular when $m_e/E \to 0$ when it is expressed in terms
of $\alpha_{\rm phys}$, because of the appearance of large
logarithms. These may be included to all orders in $\alpha
\log(E^2/m_e^2)$ simply by choosing $\mu = E$ in eq.~\pref{RGobs}
and using eq.~\pref{rgimproved} with eq.~\pref{rgbc}.

For the next section it is important that eq.~\pref{rgeqtn}
integrates so simply because the $\msbar$ renormalization is a
{\em mass-independent} scheme. That is, $d\alpha/d\mu$ depends
only on $\alpha$ and does not depend explicitly on mass scales
like $m_e$. (`On shell' renormalizations, such as where $e$ is
defined in terms of the value of a scattering amplitude at a
specific momentum transfer, furnish examples of schemes which are
not mass-independent.)

\subsubsection{Muons and the Decoupling Subtraction scheme:}

Consider now introducing a second scale into the problem by
raising the energies of interest to those above the muon mass. In
this case the underlying theory is
\eq \label{QEDwmuons}
    \Scl = - \, \nth{4} \, F_{\mu\nu} F^{\mu\nu}
    - \psibr (\Dslsh + m_e) \psi - \chibr (\Dslsh + M_\mu) \chi,
\eeq
where $\chi$ is the Dirac spinor representing the muon, and
$M_\mu$ is the muon mass. Our interest is in following how large
logarithms like $\log(M_\mu/m_e)$ appear in observables.

In this effective theory the $\msbar$ and physical electromagnetic
couplings are related to one another by:
\eq \label{newschemerelation}
    \alpha_\msbar = \left\{ 1 - {\alpha
    \over 3 \pi} \left[ \log \left( {m_e^2 \over \mu^2} \right) + \log
    \left( {M_\mu^2 \over \mu^2} \right) \right] \right\} \;
    \alpha_{\rm phys} .
\eeq
This relation replaces eq.~\pref{schemerelation} of the purely
electron-photon theory. Notice, in particular, that in this theory
the physical coupling, $\alpha_{\rm phys}$ is no longer simply
equal to $\alpha_\msbar(\mu = m_e)$ but is now equal to
$\alpha_\msbar(\mu = \sqrt{m_e M_\mu})$. The corresponding RG
equation for the running of $\alpha_\msbar$ therefore becomes:
\eq \label{newrgeqtn}
    \mu^2 {d \alpha_\msbar \over d \mu^2 } =  +
    {2 \alpha^2 \over 3 \pi} ,
\eeq
and its solution is
\eq \label{newrgimproved}
    {1 \over \alpha_\msbar(\mu)} = {1 \over
    \alpha_\msbar(\mu_0)} - {2 \over 3 \pi} \; \log\left( {\mu^2 \over
    \mu_0^2 } \right) .
\eeq

Here we see an inconvenience of the $\msbar$ renormalization
scheme: the right-hand-side of eq.~\pref{newrgeqtn} is twice as
large as its counterpart, eq.~\pref{rgeqtn}, in the pure
electron-photon theory, simply because the mass-independence of
the $\msbar$ scheme ensures that both the electron and the muon
contribute equally to the running of $\alpha_\msbar$. The problem
is that this is equally true for all $\mu$, and it even applies at
scales $\mu \ll M_\mu$, where we expect the physical influence of
the muon to decouple. Of course, the physical effects of the muon
indeed {\em do} decouple at scales well below the muon mass, it is
just that this decoupling is not manifest at intermediate steps in
any calculation performed with the $\msbar$ scheme.

The dimensionally-regularized effective lagrangian furnishes a way
to circumvent this disadvantage, by keeping the decoupling of
heavy particles manifest without giving up the benefits of a
mass-independent renormalization scheme. The remedy is to work
with minimal subtraction, but to do so only when running couplings
in an energy range between charged-particle thresholds. As the
energy falls below each charged-particle threshold, a new
effective theory is defined by `integrating out' this particle,
with the couplings in the new low-energy theory found by matching
to the coupling defined in the underlying theory above the
relevant mass scale. The scheme defined by doing so through all
particle thresholds is called the {\em decoupling subtraction}
($\dsbar$) renormalization scheme \cite{DSScheme}.

For instance, for the electrodynamics of electrons and muons, the
coupling constant as defined in the $\msbar$ and $\dsbar$ schemes
is identical for the full theory which describes energies greater
than the muon mass: $\mu > M_\mu$. For $ m_e < \mu < M_\mu$ we
integrate out the muon and construct an effective theory involving
only photons and electrons. This effective theory consists of the
usual QED lagrangian, plus an infinite number of higher-dimension
effective interactions encoding the low-energy implications of
virtual muons. Within this effective lagrangian the coupling
constant is again defined by the coefficient $Z$ in front of the
$F_{\mu\nu}F^{\mu\nu}$ term, using minimal subtraction. But,
because there is no muon within this effective theory, only the
electron contributes to its running.

The initial conditions for the RG equation at the muon mass is
obtained by matching, as in the previous sections. They are chosen
to ensure that the effective theory reproduce the same predictions
for all physical quantities as does the full theory,
order-by-order in the low-energy expansion. If there were other
charged particles in the problem, each of these could be
integrated out in a similar fashion as $\mu$ falls below the
corresponding particle threshold.

Quantitatively, to one loop the RG equation for the $\dsbar$
scheme for the theory of electrons, muons and photons becomes:
\eqa \label{dsrgeqtn}
    \mu^2 {d \alpha_\dsbar \over d \mu^2 } &=&
    {2 \alpha^2 \over 3
    \pi} , \qquad \hbox{if} \qquad \mu > M_\mu ; \nn\\
    &=&  {\alpha^2 \over 3 \pi} , \quad\quad\;\; \hbox{if} \qquad m_e < \mu
    < M_\mu ;
    \nn\\
    &=& 0,  \qquad\quad\; \hbox{if}\qquad \mu < m_e ,
\eeqa
with the boundary conditions that the $\alpha_\dsbar$ should be
continuous at $\mu = M_\mu$ and $ \alpha_\dsbar(\mu = m_e)
=\alpha_{\rm phys}$ at $\mu = m_e$. Integrating then gives
\eqa \label{finalrgresult}
    {1 \over \alpha_\dsbar(E)} &=& {1 \over
    \alpha_{\rm phys}} - {1 \over 3
    \pi} \; \log\left( {E^2 \over M_\mu^2 } \right)
    \qquad
    \hbox{for $m_e < E < M_\mu$}, \\
    &=& {1 \over \alpha_{\rm phys}} - {1 \over 3 \pi} \; \log\left(
    {M_\mu^2 \over m_e^2 } \right)  - {2 \over 3 \pi} \; \log\left(
    {E^2 \over M_\mu^2 } \right) \quad
    \hbox{for $M_\mu < E$}.\nn
\eeqa
This last expression shows how to efficiently display the various
large logarithms by running a coupling with the ease of a
mass-independent scheme, but with each particle explicitly
decoupling as $\mu$ drops through the corresponding particle
threshold.

\subsection{Power-counting examples: QCD and Gravity}

We close with a sketch of the utility of eq.~\pref{aedwdimreg} for
two important examples: the interactions amongst pions and kaons
at energies well below a GeV; and gravitational self-interactions
for macroscopic systems.

\subsubsection{Below the QCD scale: mesons}

We start with the interactions of pions and kaons at energies
below a GeV. This represents a useful low-energy limit of the
Standard Model because these mesons are Goldstone bosons for the
spontaneous breaking of an approximate symmetry of the strong
interactions \cite{CSB}. As such, their interactions are
suppressed in this low-energy limit, as a general consequence of
Goldstone's theorem \cite{Goldstone}. Because they interact so
weakly we represent them with fundamental scalar fields in the
effective theory which applies at energies $E \ll \Lambda \sim 1$
GeV \cite{Weinberg}.

The resulting scalar lagrangian has the form of eq.~\pref{leffpc},
with the constants appearing in the effective lagrangian being: $f
= \sqrt{\fpi \Lambda}$, $M = \Lambda$ and $v = \fpi$, where $\fpi
\sim 100$ MeV defines the scale of the order parameter which
describes the spontaneous breaking of the relevant approximate
symmetry. In this case the powercounting estimate of
eq.~\pref{aedwdimreg} becomes:
\eq \label{ChPTpcapp}
    A_\sse(q) \sim \fpi^2 q^2 \;
    \left( {1 \over \fpi} \right)^E \; \left( {q \over 4 \pi
    \fpi} \right)^{2L} \; \left( {q \over \Lambda} \right)^{
    \sum_{ik} (k - 2) V_{ik}} ,
\eeq
which is a famous result, due first to Weinberg \cite{WPhysica}.
The explicit suppression of all interactions by powers of
$q/\Lambda \sim q/(4\pi \fpi)$ explicitly encodes the suppression
of interactions that Goldstone's theorem requires. Since the pion
mass is $m_\pi \sim 140$ MeV, this suppression is clearly only
suppressed for scattering at energies near threshhold, $E_{\rm cm}
\sim m_\pi$.

The dominant terms in $\leff$ which govern the scattering at these
energies corresponds to choosing the smallest possible value for
which $L = 0$, and $V_{ik} \ne 0$ only if $k = 2$. Since it
happens that symmetries determine the effective couplings of all
such terms purely in terms of $\fpi$ (in the limit of massless
quarks), a great deal can be said about such low-energy meson
interactions without knowing any of the dynamical details about
their explicit wave-functions. These predictions are consequences
of the symmetry-breaking pattern, and are known as `soft pion'
theorems. Comparison of these predictions, including
next-to-leading corrections and nonzero quark masses, are in good
agreement with observations \cite{CPTReview}.

\subsubsection{General Relativity as an Effective Field Theory}

It is instructive to repeat this powercounting analysis for the
gravitational effective theory, since this case furnishes a less
familiar example. The result obtained also justifies the neglect
of quantum effects in performing practical calculations with
gravity on macroscopic scales. Even better: it permits the
systematic calculation of the leading corrections in the
semiclassical limit, should these ever be desired.

The field relevant for gravity \cite{Gravity} is the metric,
$g_{\mu\nu}$ (whose matrix inverse is denoted $g^{\mu\nu}$). For
applications on macroscopic scales we use the most general
effective lagrangian consistent with general
covariance:\footnote{No cosmological term is written here since
this precludes a perturbative expansion about flat space.}
\eq \label{GReff}
    -\Scl_{\rm eff} = \sqrt{-g} \left[ \frac12 \,\mpl^2 R + c_1 \, R^2
    + c_2 \, R_{\mu\nu} R^{\mu\nu} + c_3 R_{\mu\nu\lambda\rho}
    R^{\mu\nu\lambda\rho} + \frac{e_1}{m_e^2} R^3 + \cdots \right]\,,
\eeq
where $g = \det (g_{\mu\nu})$, while $R_{\rho\mu\lambda\nu}$,
$R_{\mu\nu} = g^{\lambda\rho} R_{\lambda\mu\rho\nu}$ and $R =
g^{\mu\nu}R_{\mu\nu}$ respectively denote the Riemann and Ricci
tensors, and the Ricci scalar, each of which involves two
derivatives of $g_{\mu\nu}$. The ellipses denote terms involving
at least six derivatives, one term of which is displayed
explicitly in eq.~\pref{GReff}. The term linear in $R$ is the
usual Einstein-Hilbert action, with $\mpl$ denoting the usual
Planck mass. The remaining effective couplings --- $c_k$ and $e_k$
--- are dimensionless, and not all of the terms written need be
independent of one another. The scale $m_e$ denotes the lightest
particle (say, the electron) to have been integrated out to obtain
this effective lagrangian.

Eq.~\pref{GReff} has the form considered earlier, with $f =
\sqrt{m_e \mpl}$, $\Lambda = m_e$ and $v = \mpl$. Furthermore,
with these choices, the dimensionless couplings of all of the
interactions {\em except} for the Einstein term, are explicitly
suppressed by the factor ${m_e^2 / \mpl^2}$. With these choices
the central powercounting result, eq.~\pref{aedwdimreg}, becomes
\cite{GravityPC,GravityPCNR}:
\eqa  \label{GRpcapp}
    \Scabr_\sse(q) &\sim& q^2 \mpl^2 \; \left( {1
    \over \mpl} \right)^E \; \left( {q \over 4 \pi \mpl}
    \right)^{2L} \; \left( {m_e^2 \over \mpl^2} \right)^{\sum_{i;k >
    2} V_{ik}}  \; \left( {q \over m_e} \right)^{\sum_{ik} (k
    - 2) V_{ik}} \nn\\
    &\sim& q^2 \mpl^2 \; \left( {1
    \over \mpl} \right)^E \; \left( {q \over 4 \pi \mpl}
    \right)^{2L} \; \left( {q^2 \over \mpl^2} \right)^{\sum_{i;k >
    2} V_{ik}}  \; \left( {q \over m_e} \right)^{\sum_{i;k>4} (k
    - 4) V_{ik}}
\eeqa
where covariance requires $V_{ik} = 0$ unless $k = 2,4,6...$, with
$k = 2$ corresponding to the Einstein-Hilbert term, $k=4$ the
curvature-squared terms, and so on.

As before, the dominant term comes from choosing $L=0$ and using
only the interactions of the usual Einstein-Hilbert action: \ie\
$V_{ik} = 0$ for $k > 2$. The dominant contribution to
gravitational physics is therefore obtained by working to tree
level with the Einstein action, which is to say that one is to
compute the classical response of the gravitational field, using
the full Einstein equations to compute this response.

The graphs responsible for the next-to-leading terms are also
simple to determine. The minimum additional suppression by
$q/\mpl$ is obtained either by working to one loop order ($L=1$)
using the Einstein action ($V_{ik}=0$ for $k>2$), or by working to
tree level ($L=0$) using precisely one insertion of one of the
curvature-squared interactions (\ie\ with $V_{i2}$ arbitrary but
$V_{i4} = 1$ for one interaction with $k=4$). Both cases give an
additional suppression of $q^2/\mpl^2$ relative to the leading
contribution. The one-loop contribution also carries the usual
additional loop factor, $(1 / 4 \pi)^{2}$.

Notice that the derivative expansion is an expansion in $q/m_e$ as
well as in $q/\mpl$, due to the inverse powers of $m_e$ which
appear in the higher-curvature terms. Notice also that all of the
$m_e$-dependence drops out for graphs constructed only using the
Einstein and the curvature-squared terms (\ie\ $V_{ik}=0$ for
$k>4$), as it should since $1/m_e^2$ first enters eq.~\pref{GReff}
at order curvature-cubed. Although the condition $q \ll m_e$ may
come as something of a surprise, it is nevertheless an excellent
expansion for macroscopic applications, such as in the solar
system.

\section*{Acknowledgements}
This review is based on a series of lectures given for the Swiss
Troisi\`eme Cycle in Lausanne, and in the University of Oslo, in
June 1995, whose organizers I thank for their kind invitations,
and whose students I thank for their questions and comments. My
research during the preparation of these lectures was funded in
part by NSERC (Canada), FCAR (Qu\'ebec) and the Killam Foundation.

\end{document}